\documentstyle[12pt,epsfig]{article}
\begin{document}
%\draft

\vskip 1cm

\title{\bf Dissipative effects on
quantum glassy systems}
\vskip 20pt

\author{
L. F. Cugliandolo$^{1,2}$, D. R. Grempel$^3$, 
\\
G. Lozano$^4$, H. Lozza$^4$ and C. A. da Silva Santos$^2$
\\
%}
%\address{
$^1$Laboratoire de Physique Th{\'e}orique de l'{\'E}cole Normale
Sup{\'e}rieure,
\\
24 rue Lhomond, 75231 Paris Cedex 05, France
\\
$^2$Laboratoire de Physique Th{\'e}orique  et Hautes {\'E}nergies, Jussieu,
\\
1er  {\'e}tage,  Tour 16, 4 Place Jussieu, 75252 Paris Cedex 05, France
\\
$^3$CEA-Saclay/SPCSI, 91191 Gif-sur-Yvette CEDEX, France
\\
$^4$ Departamento de F{\'{}}{\i}sica, FCEyN, 
Universidad de Buenos Aires,
\\
Pabell{\'o}n I, Ciudad Universitaria, 1428 Buenos Aires, Argentina
}
\date\today
\maketitle
%\widetext

\begin{abstract}
We discuss the behavior of a quantum glassy system coupled to
a bath of quantum oscillators. We show that the system 
localizes in the absence of interactions when coupled to 
a subOhmic bath. When interactions are switched on 
localization disappears and the system undergoes a phase 
transition towards a glassy phase. 
We show that the position of the critical line separating
 the disordered and the ordered phases strongly depends on
 the coupling to the bath.
For a given type of bath, the ordered 
glassy phase is favored by a stronger coupling.
Ohmic, subOhmic and superOhmic baths
 lead to different transition lines.
We draw our
conclusions from the analysis of the partition function using the
replicated imaginary-time formalism and from the study
of the real-time dynamics of the coupled system using the
Schwinger-Keldysh closed time-path formalism.
\end{abstract}

\newpage

\section{Introduction}
\setcounter{equation}{0}
\renewcommand{\theequation}{\thesection.\arabic{equation}}
\label{introduction}

 The effects of a dissipative environment
 on the dynamics of quantum  systems have been intensively
investigated
during the last two decades~\cite{review-Leggett,weiss}.
The most widely studied problem is that of a {\it single} macroscopic
variable
 coupled to a set of  microscopic degrees of freedom that act as
a bath.
The environment
is usually described in terms of its
 collective excitations
 (lattice vibrations, spin or charge fluctuations, etc) that may
 be thought of as an
ensemble of independent quantum harmonic
oscillators~\cite{CL-first,Bray-Moore,Cha,Feve,Stamp}. Their
 coupling to the system is given in terms of
  a spectral density $I(\omega) \propto \alpha\;\omega^s$ for $\omega \ll 
 \omega_c$, where $\alpha$ is a
 dimensionless coupling constant and $\omega_c$ a  high frequency cutoff.
The exponent $s$ characterizes different types of environment.
The Ohmic case ($s =
 1)$ is quite generally encountered~\cite{weiss} but
superOhmic ($s>1$) and subOhmic ($s<1$) baths
 also occur {\it e.g.} in the case of the Kondo effect in
 unconventional hosts~\cite{Fradkin,Sengupta}.

The question of how dissipation destroys quantum {\it
 coherence}~\cite{review-Leggett,Bray-Moore,Cha} in two-level
 systems ({\sc tls})
 has been extensively investigated in the literature.
The low-energy physics of many tunneling systems is well described by the
spin-boson model~\cite{review-Leggett,weiss}.  In this model,
the two equivalent degenerate states of the {\sc tls} are
 represented by the
two eigenstates $\sigma_z = \pm 1$ of an Ising
pseudo-spin. A transverse field coupled to $\sigma_x$ (say)
 represents the tunneling matrix element.
Much is known about the properties of this model and
 its relationship to several other models including the 1-D
 Ising model with inverse squared
 interactions~\cite{Charu}, the anisotropic Kondo model~\cite{Andyu,Costi}
or the resonant model~\cite{Mura}. Three different
 regimes are possible depending on the value
 of $\alpha$: in the Ohmic case, at zero temperature, there is a
phase transition at $\alpha=1$~\cite{Bray-Moore,Cha}.
For $\alpha<1$  there is tunneling and two distinct regimes
develop. If
$\alpha<1/2$ the system relaxes with damped coherent oscillations;
in the intermediate region $1/2<\alpha<1$ the system relaxes incoherently.
For $\alpha>1$ quantum tunneling
is suppressed and $\langle \sigma_z\rangle \neq 0$ signalling that the
system remains localized in the state in which it was prepared.

These results also hold for sub-Ohmic baths
while weakly damped oscillations persist for super-Ohmic
baths~\cite{review-Leggett}.
At finite temperatures (but low enough such that
thermal activation can be neglected),
there is no localization but the probability of finding the system
in the state it was prepared decreases slowly with time 
for $\alpha>\alpha^{\sc crit}$. 

These conclusions, derived for a {\it single} {\sc tls}
 interacting with a bath, can be applied to  a macroscopic
  system in the {\it diluted} regime, {\it i.e.}
when the interactions between the {\sc tls} are
 unimportant compared with those between a {\sc tls} and the bath~\cite{Gozico}. 
There are, however,
 physical systems that can be viewed as a
{\it dense} set of {\sc tls} in which
their mutual interactions can no
 longer be neglected. 
The question then arises as to which are the effects of the interplay
 between the interactions between the {\sc tls} and their coupling to the noise
 on the  physics of the interacting system.

In this paper we discuss this
 issue in the context of a {\it glassy} macroscopic system
with {\it random}, {\it long-ranged} interactions. This situation is
 realized
 experimentally in systems such as  uniaxial spin glasses
in a transverse magnetic field~\cite{aeppli}
and  disordered Kondo alloys~\cite{lohn,tabata}.  Metallic glasses
 with tunneling defects are also systems in which the
 effects that are of interest here could be observed experimentally.

In thermodynamic equilibrium, in the absence of the bath, the
interactions between the {\sc tls} lead to the appearance of an
ordered state at low enough temperature. If the 
interactions are of random sign, as in the models we consider here,
the latter will be a spin glass  ({\sc sg}) state. In this phase 
the symmetry between the states $\sigma_i^z = \pm 1$
 at any particular site is broken but there is no global
magnetization, $\sum_i \langle \sigma_i^z \rangle = 0$. 
Since the coupling to the bath also tends to
 locally break the symmetry between the degenerate states of the
{\sc tls}, both interactions compete with the tunneling term in
the Hamiltonian. We thus expect the presence of noise to increase the
stability of the {\sc sg} state against quantum fluctuations. The
consequences of this fact are  
particularly interesting when the coupling to the bath
leads by itself to localization at some $\alpha=\alpha^{\sc crit}$. 
 Consider a system of size $N$ with $\alpha > \alpha^{\sc crit}$ at $T=0$
and suppose that we turn off the interactions between the {\sc tls}.  
The ground state of the system is then 
$2^N$-fold degenerate as each {\sc tls} can be in one of
the states $\langle \sigma_i^z \rangle = \pm \;\sigma_0$ (say)
independently. If we
now turn on an infinitesimal random interaction between the {\sc tls},
 this macroscopic degeneracy will be immediately lifted as the system
will select among its 
$2^N$ degenerate configurations the one (or one among the ones) 
that minimizes the interaction
energy. If we denote by $\tilde{J}$ the typical scale of the
interactions and by $\alpha^{\sc crit}$ the localization threshold, we
thus expect a quantum critical point at $\tilde{J}=0$,
$\alpha=\alpha^{\sc crit}$ between a quantum paramagnet and the ordered
state such that, for  $\alpha > \alpha^{\sc crit}$, the {\sc
sg} phase survives down to $\tilde{J}=0$. 

A system of non-interacting localized {\sc tls} and a
{\sc sg} state {\it in equilibrium} are in some way similar: in both cases
$\sum_i \langle  \sigma_i^z \rangle =0 $ and the presence of order is
reflected by a non-vanishing value of the long-time limit of the
correlation function, 
$q_{\sc ea} = \lim_{t \to \infty}
N^{-1} \sum_i \langle  \sigma_i^z(t) \sigma_i^z(0)
\rangle $ (since we assume equilibration the correlation is stationary and 
the reference time can be taken to be zero).
However, this resemblance is only superficial. In the
renormalization-group language, $\tilde{J}$ is a relevant
variable~\cite{quimiao}. 
Therefore the details of the
dynamics of the two systems are expected to be quite different, in
particular the way in which the correlation function reaches its
asymptotic limit,  $q_{\sc ea}$, that determines the low-energy part
of the excitation spectrum of the system. 

Further differences between the  localized state and 
the {\sc sg} state are 
seen from the study of the out of equilibrium relaxation of such states.
Indeed, an important feature of glassy systems is that 
their low-temperature dynamics occurs out of equilibrium. If the 
system is macroscopic, its size $N$ is very large (diverges
in the thermodynamic limit). In a realistic macroscopic situation, 
the asymptotic long-time limit follows this large size limit.  
Many experiments, simulations and analytical studies show that 
the time needed to reach equilibrium after entering the glassy
phase diverges so quickly that the relevant relaxation 
occurs out of equilibrium. The dynamics at low temperatures is then 
non stationary, {\it i.e.}
the dynamic correlation functions loose time translation invariance. 
If $t_w$ denotes the time elapsed
since a quench from the high temperature phase into the {\sc sg}
phase, $C(t + t_w,t_w)$ depends on both $t$ and $t_w$. The order
in which the limits $t_w\to\infty$ and $t\to\infty$ are taken is in this case
very important. For sufficiently long $t$ and $t_w$
but in the regime $t \ll t_w$, the dynamics is
stationary and the correlation function reaches a plateau  $q_{\sc
ea}$. Much of what was said above for the equilibrium state also holds for
this stationary regime. However, for times $t > t_w$, the system enters an
{\it aging} regime where the correlation function depends on 
the waiting-time $t_w$ explicitly.  In this regime,
the dynamic correlation function vanishes at long times, $\lim_{t
\to \infty} C(t + t_w,t_w) = 0$, at a rate that depends on $t_w$.
(This is to be confronted to the dynamics 
in the localized state, where $C(t+t_w,t_w)$ reaches, for any 
waiting-time $t_w$ and long enough $t$ a plateau that it never 
leaves.) In this regime, even for $\alpha > \alpha^{\rm crit}$, small
interactions will result in the  {\it destruction} of 
localization of the {\sc tls} at long enough times.

The problem of a single {\sc tls} being a difficult one,
that of an infinite set of interacting {\sc tls} seems hardly
solvable at this stage. Therefore, as a first step,
we shall focus on the low-temperature dynamics of a very simple
model that mimics some of the features of more realistic ones. This is
  a quantum
generalization~\cite{Gold,Culo,Cugrsa1,Cugrsa2,Bicu} of the random  
 $p$-spin spherical model~\cite{crisanti} coupled to a
 bath of quantum harmonic oscillators. The principal merit of this
model is that it is simple enough that it can be studied in
detail. Yet, many of its properties are generic and  expected to hold
at least qualitatively for more realistic models~\cite{young}.  
The usual methods of
equilibrium quantum statistical mechanics are inappropriate to
describe the nonstationary situation. We solve the model using two methods
especially designed to treat systems out of equilibrium. 

One is
based on the Schwinger-Keldysh real-time approach to
non-stationary systems. It was first applied in this context to
the quantum $p$-spin model in Ref.~\cite{Culo} and used
subsequently in other cases including 
the SU(${\cal N}$) fully connected
Heisenberg model in the limit of large ${\cal N}$~\cite{Bipa} and
the soft spin version of the Sherrington - Kirkpatrick model
\cite{Chamon}. It allows one to obtain the full time dependence 
of the correlation $C(t+t_w,t_w)$. 

The second method is based on the {\it Ansatz} of marginal
stability ({\sc ams}) within the replica analysis of the 
partition function.  Originally developed  for classical
systems~\cite{KT}, this method
was recently used to discuss the low-temperature
properties of quantum glassy
systems~\cite{Cugrsa2,ledou,antoine}. Its main
advantage
is that it uses a formalism that is
 closely related to the imaginary-time approach to
 equilibrium quantum statistical mechanics.
In Ref.~\cite{Cugrsa2} the {\sc ams} was extensively applied
to the quantum spherical $p$-spin model in the absence of the bath. It
was shown that the position of the dynamic transition line predicted
by this method coincides precisely with that obtained using the
real-time approach. It was also shown that the time dependent
correlation function computed
using the {\sc ams} in the absence of the bath
is identical to the {\it stationary} part of the non-equilibrium
correlation function ($C>q_{\sc ea}$) 
when one takes the long-time limit first and
the limit in which the coupling to the bath goes to zero next.
The marginality condition imposed by the {\it Ansatz} is intimately
related to the fact that the correlation will further decay 
from $q_{\sc ea}$ towards zero. (The details of this second 
decay as, for instance, the two-time scaling are not accessible with 
this method.) A localized solution with $C(t+t_w,t_w)$ approaching, and
never leaving, the plateau at $q_{\sc ea}$ corresponds, in 
replica terms, to a stable replica symmetry solution. 
 In this paper we extend the {\sc ams} to study the
dynamics of the model in the case in which the system is coupled to
the environment.

This paper is organized as follows. In  Section~\ref{themodel} we
motivate and introduce our model and discuss its relationship to the
more usual spin-boson model.
In Section~3 we outline the imaginary-time formalism
used to solve the problem at equilibrium and within the {\sc ams}.
We compute the partition function of the coupled system and
determine its phase diagram in both situations. We also discuss
the long-time dynamics of the coupled system using a very accurate
long-time approximation~\cite{Cugrsa2,Bicu} that allows us to
solve the model analytically. This approximation is then
used to discuss
 the influence of a coupling to different types of environment on the
$T=0$ quantum phase transition.
 The real-time dynamic of the system is discussed in Section~\ref{real-time2}
and the results are compared with those of
Section~3. Section~\ref{conclusions} contains  a brief 
summary of our main results and our
concluding remarks.

\section{The model}
\setcounter{equation}{0}
\renewcommand{\theequation}{\thesection.\arabic{equation}}
\label{themodel}

In order to motivate our model, we start by considering a collection of $N$
identical interacting {\sc tls} coupled  to a bath of independent
harmonic oscillators~\cite{Feve,caldeira}. We assume for the moment
and until otherwise stated
that the {\it combined} system is in thermodynamic equilibrium.
The Hamiltonian of the coupled system may be written as  
\begin{equation}
\label{tls1}
H=H_S + H_B + H_{SB}\;,
\end{equation}
where $H_S$, $H_B$, and $H_{SB}$ denote the Hamiltonians of the
system, the bath and their coupling, respectively. These are
given by:
\begin{eqnarray}\label{hamis}
H_S&=& - \Delta \sum_{i=1}^N \sigma_i^x +
V(\sigma_1^z,\sigma_2^z,\cdots,\sigma_N^z)\;,\\
\label{hamib}
H_B& =& {1\over 2}\sum_{\ell}\left({p_{\ell}^2\over m_{\ell}}+
m_{\ell}\omega_{\ell}^2
x_{\ell}^2\right)\;, \\
\label{hamisb}
H_{SB}& =& -\sum_{i,\ell} c_{\ell}^i\, x_{\ell} \sigma_{i}^z
\;.
\end{eqnarray}
Here, the Pauli-matrices $\sigma^\mu _i$ represent the {\sc tls}'s pseudo-spins, $\Delta/\hbar$ is their tunneling frequency and $V$ their
mutual interaction potential that we leave unspecified for the moment.
 $x_{\ell}$ and $p_{\ell}$ are the coordinate and momentum of
 the $\ell$-th oscillator  and  $m_{\ell}$ and  $\omega_{\ell}$ its mass and
frequency, respectively. We denote by $c_{\ell}^i$
the coupling constant between the $i$-th {\sc tls} and
the $\ell$-th oscillator.

Using standard methods~\cite{grabert,caldeira} the oscillator degrees
of freedom may be integrated out to express the
 partition function of the system solely in terms of the {\sc tls} variables as:
\begin{equation}
Z\equiv {\rm Tr}\, e^{-\beta\hat{H}}=\,
\mathop{\rm Tr}\limits_{\{\sigma\}}\;
\left[\;{\cal T}\;\exp\left(-{S\over \hbar}
\right)\right]
%\; ,
\label{partTLS}
\end{equation}
with
\begin{equation}
\label{actionTLS}
S=\int_0^{\hbar \beta} d\tau \left\{-\Delta \sum_i \sigma^x_i(\tau) +
V[\vec{\sigma}^{\;z}(\tau)] \right\}
+ {1\over 2} \sum_{i j}\;\int_{0}^{\hbar\beta}\int_{0}^{\hbar\beta}d\tau\, d\tau'\,
K_{i j}(\tau - \tau') \; \sigma_i^{\;z}(\tau) \sigma_j^{\;z}(\tau')
\; ,
\end{equation}
 where ${\cal T}$ the imaginary-time ordering operator and we have introduced the notation $\vec{\sigma}^{\;z} =
\left(\sigma_1^z,\sigma_2^z,\cdots,\sigma_N^z\right)$.

The
kernel $K_{i j}(\tau)$ in Eq.~(\ref{actionTLS}) is \cite{grabert}
\begin{equation}
K_{i j}(\tau)={1\over \hbar\beta}\sum_{\omega_k}
 \tilde{K}_{i j}(\omega_k) \exp (-i\omega_k\tau)
\; ,
\label{kernel}
\end{equation}
where $\omega_k = 2\pi k/(\hbar\beta)$ are the Matsubara frequencies,
the coefficients $\tilde{K}_{i j}(\omega_k)$ are given by
\begin{equation}
\tilde{K}_{i j}(\omega_k) \equiv  \int_0^{\infty}{d\omega \over \pi}
{I_{i j}(\omega)\over
\omega}{ \omega_k^2 \over \omega^2 + \omega_k^2}
\; ,
\label{qui}
\end{equation}
and we have introduced the spectral density of the
environment $I_{i j}(\omega)$ through
\begin{equation}
I_{i j}(\omega)=\pi \sum_{\ell}{(c_\ell^i {c^j}^{\star}_{\ell}+ 
{c^i}^{\star}_\ell {c^j}_{\ell})
\over 2 m_{\ell}\omega_{\ell}}\delta(\omega
-\omega_{\ell})
\; .
\label{spectral1}
\end{equation}
We make the
simplifying assumption that the
dynamic interaction between {\it different} {\sc tls}
 generated by integration over  the degrees of freedom of the bath can
be
neglected compared to the static interaction potential included in
$V[\vec{\sigma}^{\;z}]$. Therefore we write
\begin{equation}
\label{chantada}
I_{i j}(\omega) = \delta_{i j} I(\omega)\;,
\end{equation}
and we choose the standard parametrization~\cite{weiss}
\begin{eqnarray}
I(\omega) = 2 \alpha \hbar
\left(\frac{\omega}{\omega_{ph}}\right)^{s-1} \omega
\;e^{-\omega/\omega_c} \; ,
\label{Iomega}
\end{eqnarray}
where $\alpha$ is a dimensionless coupling constant, $\omega_c$ is
a high frequency cutoff and $\omega_{ph}$ is a microscopic phonon
frequency necessary in the non-Ohmic cases in order to keep
$\alpha$ dimensionless. For simplicity, we shall restrict
 the exponent $s$ to lie in the interval $0 < s < 2$. With this choice
the integral on the right-hand side of Eq.~(\ref{qui})
converges without the need of
introducing an infrared cutoff and the upper cutoff may be eliminated
by taking the limit $\omega_c \to \infty$. This leads to
the expression:
\begin{equation}
\label{quiw}
\tilde{K}(\omega_k)={\alpha \hbar \over \omega_{ph}^{s-1}
\sin\left(\pi s /2\right)}\; \left| \omega_k
\right|^s\;.
\end{equation}

We shall consider diagonal $p$-spin interactions of the form
\begin{equation}
\label{defV}
V[\vec{\sigma}^{\;z}]= \sum^{N}_{i_1<...<i_p} J_{i_1...i_p}
\sigma_{i_1}^z ... \sigma_{i_p}^z \; ,
\end{equation}
with random couplings $J_{i_1...i_p}$. These are taken from a Gaussian
distribution  with zero mean and variance
\begin{equation}
\label{corrJ}
\overline{\left(J_{i_1...i_p}\right)^2} = {{\tilde{J}}^2
p!/(2N^{p-1})}\;,
\end{equation}
where the overline represents an average over disorder.

In the case $p=2$ and for an Ohmic bath ($s=1$),
Eq.~(\ref{actionTLS}) is equivalent to the action of
a disordered Kondo alloy model~\cite{metallic}. In this context~\cite{Costi}, 
$\Delta = J_{\perp}^{\sc k}$, the transverse Kondo coupling 
and $(1 - \alpha) \ll 1$ is
proportional to $J_{||}^{\sc k}$ the parallel Kondo coupling. In
the opposite limit, $\alpha \ll 1$, Eq.~(\ref{actionTLS})
is a representation of the partition
function of the SK spin-glass model in a transverse magnetic
field~\cite{SK-transv} weakly coupled to a phonon (or
spin~\cite{Stamp}) bath.

The main difficulty in solving the quantum statistical
mechanical problem defined by Eq.~(\ref{actionTLS}) stems from
the discrete nature of the spins. It was shown in
Ref.~\cite{Cugrsa2} (hereafter referred to as CGS) that, in the
absence of the bath, a solvable (and yet non-trivial) model can be
obtained by generalizing the $\sigma^z$ eigenvalues  $s_i=\pm 1$
 to continuous variables $-\infty <s_i < \infty $, and replacing
 the hard constraint $s_i^2 = 1$
by the soft spherical constraint $\sum_i \langle s_i^2\rangle =
N$. The derivation of the effective continuous model in the
presence of the bath is analogous to that given in CGS for the
isolated system 
and we refer the reader to this reference 
 for the details. After
performing a Trotter-like decomposition of the time-ordered
exponential in Eq.~(\ref{partTLS}) the trace becomes a functional
integral over classical fields $s_i(\tau)$ and the tunneling term
in the action acquires the low-energy form
\begin{equation}
\label{low-E}
 - \Delta \int_0^{\hbar \beta} d\tau \sigma^x(\tau)
\rightarrow
 \frac{M}{2}\;\int_0^{\hbar \beta}\;d\tau \left(\frac{\partial
s^z}{\partial \tau}\right)^2\;,
\end{equation}
where we introduced the {\it mass}
\begin{equation}\label{mass}
M = \frac{\hbar \tau_0}{2}\;\ln\left(\frac{\hbar}{\Delta
\tau_0}\right)\;,
\end{equation} 
and $\tau_0$ is a  cutoff representing a microscopic spin-flip time
that we identify with $\omega_c^{-1}$. Since we work in the regime in
which
$\hbar \omega_c$ is the highest energy scale in the problem, $0 < M < \infty$.

The continuous version of Eq.~(\ref{actionTLS}) is thus
given by
\begin{eqnarray}
S&=& {1 \over 2} \sum_i\, \int_{0}^{\hbar\beta}\, d\tau \left[ M
\left({\partial s_i(\tau)\over \partial\tau}\right)^2
 + \int_{0}^{\hbar\beta}d\tau'\,
K(\tau - \tau') \;s_i(\tau) s_i(\tau')  +
{z\over 2}
(s_i^2(\tau) - 1) \right]\,\nonumber\\
& & - \sum^{N}_{i_1<...<i_p}\,\int_{0}^{\hbar\beta}d\tau\, J_{i_1...i_p}\,
s_{i_1}(\tau)\dots s_{i_p}(\tau)
\; ,
\label{action}
\end{eqnarray}
where $z$ is a Lagrange multiplier that enforces the spherical
constraint
\begin{equation}
\label{constraint}
\langle
\vec{s}(\tau) \cdot    \vec{s}(\tau')
\rangle\left|_{\tau=\tau'}\right. = {1\over \hbar \beta} \sum_k
\langle \left| \vec{s}(\omega_k) \right|^2\rangle = N\;,
\end{equation}
where the angular brackets represent the average with respect to the
action (\ref{action}). 
Equations~(\ref{action}) and (\ref{constraint}) define
 the quantum $p$-spin spherical model that we 
discuss in the rest of this paper. The mass parameter $M$ is a measure of the
 strength of quantum tunneling. If $\Delta \tau_0/\hbar \ll 1$,  $M$
 is large. In this case, the gradient term favors
 configurations in which $s_i(\tau)$ is almost $\tau$-independent. The
 partition function is then largely dominated by the contribution from
 the static fluctuations ({\it i.e.} those with $\omega_k=0$).
Since $\tilde{K}(\omega_k=0)= 0$,
 these variables are unaffected by the coupling to the bath which
 drops out of the partition function  in
 this limit. With increasing $\Delta$, $M$ decreases and the amplitude of
 the quantum fluctuations becomes large. The $\tau$-dependence of $s_i(\tau)$
  then becomes essential.
 
There are three points worth discussing before
presenting the solution of the model. The first one is their dependence on
the value of $p$. For $p=2$ the action is quadratic and the
problem is readily diagonalizable by Fourier transformation. This
simple case was extensively discussed in the literature both
without~\cite{p2} and with~\cite{metallic} a bath. 
In the former case, the
competition between the mass and interaction terms in
Eq.~(\ref{action}) leads to the existence of  a critical mass  $M_c \sim \hbar^2/\tilde{J}$ above which
the ground state of the
system overcomes quantum fluctuations and acquires glassy
order.
 For $p=2$, however, the ordered phase  is of a trivial type, with a
structureless order parameter (see below). Its physical
properties are
non-generic and qualitatively different from those of discrete
spin systems. The presence of a coupling to the bath does not change
this
 situation.
For all $p > 2$ the ordered ground state is 
non-trivial~\cite{Gold,Culo,Cugrsa1,Cugrsa2,Bicu} 
and the model shares
a number of qualitative features with  more realistic ones. 
Therefore, from here on, we shall 
discuss this case, choosing the particular value $p=3$ in our numerical
calculations.

The second point is about the case $\tilde{J}=0$.  In this case 
 Eq.~(\ref{action}) reduces to a
simplified model for a {\sc tls} whose physics differs in some 
 ways from that of real two-level systems. For $\tilde{J} = 0$ we have
\begin{equation}
\label{tls} \langle \left|s(\omega_k) \right|^2
\rangle_{\tilde{J}=0} = {\hbar \over M \omega_k^2 +  z + 
\tilde{K}(\omega_k)}\;,
\end{equation}
where $\tilde{K}(\omega_k)$ is defined in Eq.~(\ref{quiw}) 
and Eq.~(\ref{constraint}) at $T=0$ reads
\begin{equation}
\label{const_tls} 1 = {1 \over \pi} \int_0^\infty d\omega {\hbar
\over M \omega^2 + \tilde{K}(\omega) + z} \equiv f_s(z)\;.
\end{equation}
We consider the Ohmic case first. For $s=1$, Eq.~(\ref{tls}) is the
propagator of a simple damped harmonic oscillator with frequency
$\omega_0 = \sqrt{z/M}$ self-consistently determined by
Eq.~(\ref{const_tls}). From the position of the poles 
of Eq.~(\ref{tls}) we see that there is a transition between
 underdamped and overdamped regimes at $z=\alpha^2
\hbar^2/(4 M)$. Using this value of $z$ in Eq.~(\ref{const_tls}) (with
$s = 1$)
and solving for $\alpha$ we find that this occurs at
$\alpha = 2/\pi$, independent of $M$. Away
from this value we easily
  find the following limiting behaviors:
\begin{eqnarray}
\label{tls_limit}
z = \left\{
\begin{array}{ccc}
{\hbar^2 \over 4 M}\;\left(1 - { 4 \alpha \over \pi}+ \cdots \right)\;, & \alpha
\ll 1 \\
\\
{ \hbar^2 \alpha^2 \over M}\;\exp\left( - \pi \alpha\right)\;,  & \alpha
\gg 1
\end{array}
\right.\;.
\end{eqnarray}
For $\alpha \ll 1$ the system exhibits weakly damped oscillations
with frequency $\omega_0 \sim \hbar/M$. In the opposite limit,
$\alpha \gg 1$, the correlation function decays exponentially with
a time-constant that increases exponentially with the strength of
the coupling, $\tau \sim \alpha^{-2}
 \exp\left( \pi \alpha\right)$.

Comparing with the results for the spin-boson model summarized in
the Introduction, we see that the transition between coherent and
incoherent motion at a universal value of $\alpha < 1$ is
preserved in the spherical model but the localization transition
at $\alpha = 1$ is replaced by 
 a crossover
to a high coupling regime characterized by an exponentially small
energy scale $\propto \exp\left( - \pi \alpha\right)$. In this
regime, tunneling is not suppressed but its rate is strongly
reduced.

In the superOhmic case there is no localization transition either.
One can easily show that for $s > 1$ the expression on the second
line of Eq.~(\ref{tls_limit})  is replaced by  $z \sim \hbar
\omega_{ph} \alpha^{1/(1 - s)}$ for $\alpha \gg 1$. The decay rate
of the correlation function still decreases continuously as the
strength of the coupling to the bath increases but only as a power
law.

In the case of a subOhmic environment the situation is different.
For $s < 1$ the integral on the right-hand side of
Eq.~(\ref{const_tls}) is finite at $z =0$ where  it takes
its maximum value.
For $f_s(0) < 1$ Eq.~(\ref{const_tls}) cannot be satisfied for any
positive value of $z$. This phenomenon, completely analogous to
Bose-Einstein condensation, signals a localization transition. Solving
 the equation $f_s(0) = 1$ we find the critical coupling given by
\begin{equation}
\label{loc_tra}
\alpha^{\rm crit} \sim  \left(\hbar \over M \omega_{ph}\right)^{1-s}\;.
\end{equation}
Conversely, for any value of $\alpha$ the system localizes for a
sufficiently high value of $M$. These results are analogous to those
obtained for discrete {\sc tls}~\cite{review-Leggett}.   

One must keep these differences between the original model and its
spherical version in mind when interpreting our results, especially 
those discussed in Section~\ref{qpt}.
In Section~4 we shall illustrate the interplay between 
the localization observed for $\tilde J=0$ and  
and the glassy dynamics that appears when $\tilde J>0$.

The third point we want to stress is that
the coupled system can be thought as describing the motion of a
quantum Brownian particle of mass $M$,
constrained to move on a $N$-hypersphere
of radius $\sqrt{N}$, in the presence of a random potential $V(\vec{s})$.
The Brownian nature of the motion arises because of its interaction
with the quantum thermal bath. The infinite dimensional spherical limit 
yields, however, unphysical results if one wants to compare it to 
the well-known problem of the 
diffusion of a free quantum particle coupled to a phonon bath 
in a finite $D$ dimensional space. While here we find a localization 
transition when the bath is subOhmic, such a transition does not exist in 
the absence of disorder in the finite dimensional problem. 

\section{Replica solution}

It is by now well established that several properties of 
disordered systems can be derived with the help of the 
replica trick. This approach enables one to derive an effective 
action for an imaginary time dependent matrix order parameter.
It has been noticed that different {\it Ansatze} that 
parametrize this order parameter describe different 
physical situations as thermal equilibrium (equilibrium 
condition) and the asymptotic dynamic regime ({\it Ansatz} of 
marginal stability -- {\sc ams}). The bulk of this Section is devoted
to the analysis of the consequences of the {\sc ams}. 
The definition of this {\it Ansatz} excludes localization 
as a possibility. As discussed in the Introduction we expect 
that as soon as interactions are switched on, full localization 
is replaced by a glassy solution with non-trivial dynamics. 
This argument justifies the use of the {\sc ams} from the start.   
We briefly comment at the end on the equilibrium properties of 
the model.

\subsection{Formalism}
\setcounter{equation}{0}
\renewcommand{\theequation}{\thesubsection.\arabic{equation}}
\label{formalism}

The presence of disorder makes it necessary to compute the averages of
all physical quantities and, in particular, of the free-energy. To this
effect we use the replica trick, {\it i.e.} we write
\begin{equation}
\beta \overline f=-{1\over N}\overline{\ln Z}=
-{1\over N}\lim_{n_\to 0}{1\over n}\ln \overline{Z^n}
\; .
\end{equation}
The derivation of the expression for the free-energy associated to
the action in Eq.~(\ref{action}) closely follows that performed
for the isolated system in CGS where the interested reader
will find all the necessary details. An imaginary-time
dependent order parameter $Q_{ab}(\tau,\tau')$ is defined as
\begin{equation}
Q_{ab}(\tau,\tau') = \frac{1}{N}
\overline{\langle {\vec s}_a(\tau) \cdot    {\vec s}_b(\tau') \rangle }
\; ,
\end{equation}
where $a,b$ are replica indices. The spherical constraint imposes
the restriction $Q_{aa}(0)=1$. We are interested in a
stationary situation in which $Q_{ab}(\tau,\tau')$ depends only on time
differences and is a periodic function of its argument with period
$\beta \hbar$. We thus introduce the Fourier transforms
$\tilde{Q}_{ab}(\omega_k) = \int_0^{\hbar \beta}\;d\tau Q_{ab}(\tau)
\exp(i\omega_k)$ in terms of which the averaged free energy is found as
\begin{equation}
\label{betaf}
\beta \overline{f} = \lim_{n \to 0} G_0\;,
\end{equation}
where
\begin{eqnarray}
2G_0
& = &
- \frac{1}{n} \sum_k \mbox{Tr} \ln
\left[ (\beta\hbar)^{-1} {\tilde{\bf Q}} \right]
-\sum_k\left(1-\frac{i}{n\hbar}\sum_{ab}{\tilde{O}}_{ab}
(\omega_k){\tilde{Q}}_{ab}(\omega_k)\right)
\nonumber\\
& &
-{{\tilde{J}}^2 \beta\over 2\hbar n}\sum_{ab}
\int_0^{\beta\hbar} d\tau \left(
{1\over \hbar\beta}\sum_k\exp(-i\omega_k\tau)
{\tilde{Q}}_{ab}(\omega_k)\right)^p
- \beta z
\; ,
\label{eq:go}
\end{eqnarray}
and the operator $\tilde{O}_{ab}(\omega_k)$ is defined by
\begin{equation}
\tilde{O}_{ab}(\omega_k)\equiv
-i\delta_{ab}\left(M\omega_k^2 + z + \tilde{K}(\omega_k)\right)
\; .
\label{eq:op}
\end{equation}

The equations of motion are found from the
 saddle-point of the free-energy
with
respect to  variations of  $\tilde
Q_{ab}(\omega_k)$. They read:
\begin{equation}
{1\over \hbar}(M\omega^2_k + z
+\tilde K(\omega_k))\delta_{ab} = \left(\tilde{\bf Q}^{-1}\right)
_{ab}(\omega_k) + {{\tilde{J}}^2 p\over 2\hbar^2}
\int_0^{\hbar\beta}d\tau\exp(i\omega_k\tau)Q^{p-1}_{ab}(\tau)
\; .
\label{eq:ft}
\end{equation}
Equation~(\ref{eq:ft})  together with the spherical constraint
$Q_{aa}(\tau=0) = 1$
determine the
different phases in the model.

In the following, we discuss the solutions to Eq.~(\ref{eq:ft}).
Except when otherwise stated, we shall work with
dimensionless variables. These are defined by measuring energies in units
of $\tilde{J}$ and time in units of $\hbar/\tilde{J}$. The strength of
quantum tunneling and of the
coupling to the bath
are then measured by the parameters
\begin{equation}
\label{units}
\Gamma\equiv \frac{\hbar^2}{(\tilde{J}M)}
\;\;,\;\;\;\;\;\;\;\alpha_s \equiv {\alpha \over \sin\left(\pi
s/2\right)} {\left(\hbar
\omega_{ph} \over \tilde{J}\right)^{1-s} }
\;,
\end{equation}
 respectively.

In the paramagnetic phase
({\sc pm})
$Q_{ab}(\omega_k)$ is a diagonal matrix,
\begin{equation}
\tilde{Q}_{ab}(\omega_k)=\tilde{q}_{d}(\omega_k)\delta_{ab}
\; .
\end{equation}
Equation~(\ref{eq:ft}) then reduces to
\begin{equation}
{\omega^2_k \over \Gamma}  + z
+\alpha_s |\omega_k|^s = {1 \over \tilde{q}_d(\omega_k)}
+ { p\over 2}
\int_0^{\beta}d\tau\exp(i\omega_k\tau)q_d^{p-1}(\tau)
\; .
\label{eq:pm}
\end{equation}
In the {\sc sg} phase,
we search for 1-step replica symmetry breaking ({\sc rsb})
solutions of the form
\begin{equation}
\tilde{Q}_{ab}(\omega_k)=(\tilde{q}_{d}(\omega_k)- q_{\sc ea})
+ q_{\sc ea} \epsilon_{ab} ,
\end{equation}
where $\epsilon_{ab}=1$ if $a$ and $b$ belong to the same diagonal
block of size $m\times m$ and zero otherwise, and we introduced
the Edwards-Anderson order parameter $q_{\sc ea}$. It was shown in
CGS that this
 {\it Ansatz} is an {\it exact} solution of the isolated model. The
proof still holds in the presence of the bath provided that
 $\lim_{\omega \to 0} \tilde{K}(\omega) = 0$, which is verified
 here (cf. Eq.~(\ref{quiw})).

To completely determine the order-parameter matrix, $q_{\sc EA}$
and $m$ must be computed. As discussed in detail in CGS, this may
be done in two different ways, each leading to a physically
different state.  Within the {\it Ansatz of marginal stability}, 
$q_{\sc EA}$ is
determined by  extremization of the free energy, $m$ is chosen such
that the stability of the ordered state is marginal, {\it i.e.}, that
its excitation spectrum contains a zero-energy mode.

 Decomposing the diagonal
order-parameter $\tilde{q}_d(\omega)$ in a singular and a regular
part,
\begin{equation}
\label{regular}
\tilde q_d(\omega_k)=\beta q_{\sc ea}\delta_{\omega_k}
+
\tilde q_{\sc reg}(\omega_k)
\; ,
\end{equation}
an equation for $\tilde q_{\sc reg}(\omega_k)$ can be derived by a
straightforward generalization of the results of CGS to the case
in which noise is present. It reads:
\begin{equation}
\label{chireg}
\left[ \frac{\omega_k^2}{\Gamma}
+ z'
+ \alpha_s |\omega_k|^s -\left(\tilde \Sigma_{\sc reg}(\omega_k)-
\tilde \Sigma_{\sc reg}(0)\right)\right]\tilde q_{\sc reg}(\omega_k)=1
\; ,
\end{equation}
with
\begin{eqnarray}
\label{parameters}
z'&= &\frac{p}{2}\, \beta m \, q_{\sc ea}^{p-1}
\frac{1+x_p}{x_p}\;\,\\
\beta m &=& (p - 2) \sqrt{\frac{2}{p (p - 1)}}
q_{\sc ea}^{-p/2}\; ,\\
\beta {q_{EA}\over \tilde{q}_d(0)}& =& {x_p \over m + x_p}\;,
\end{eqnarray}
and
\begin{equation}
\label{sigma}
\tilde \Sigma_{\sc reg}(\omega_k)=\frac{p}{2}\int_0^{\beta} d\tau
\left(q^{p-1}(\tau)-q_{\sc ea}^{p-1}\right) \cos\left(\omega_k\tau\right).
\end{equation}
The parameter $x_p$ takes the  value (see CGS for the details of
the derivation),
\begin{equation}
\label{xpmar}
x_p = p - 2\;.
\end{equation}

For given $p$ and $\alpha$,
Eqs.~(\ref{chireg})-(\ref{sigma}) have solutions with $q_{\sc ea} \ne 0$
only for low enough values of $\Gamma$ and $T$. Otherwise, thermal or
quantum fluctuations destroy the ordered state. There is
thus  a
boundary $\Gamma_{\rm c}(T)$ in the $T - \Gamma$ above which the
system is in the {\sc pm} state.  We determine its shape below.

\subsection{The dynamic phase diagram}
\label{sec:results}
\setcounter{equation}{0}
\renewcommand{\theequation}{\thesubsection.\arabic{equation}}

We determined the phase diagram for the
coupled system for $p=3$ using the numerical methods described in
CGS. A critical line with a second order section (close to the 
classical critical point $(T_d, \Gamma=0)$) 
and a first order section (close to the 
quantum critical point $(T=0, \Gamma_d)$) is also obtained in the
presence of an environment. The second order critical line is 
determined by the condition $m=1$, the first order critical line
is defined 
as the locus of the points where a marginally stable
solution first appears with decreasing $\Gamma$ for $T$ fixed 
(see Fig.~\ref{m-Gamma}). 
For each $\Gamma$ and $\alpha$ this defines a 
{\it dynamic}  transition temperature $T_d(\Gamma,\alpha)$. It was
shown in CGS that  $T_d(\Gamma,\alpha)$ precisely coincides with the
temperature below which the real-time dynamics of the system looses
time-translation invariance and the fluctuation-dissipation theorem
({\sc fdt}) is violated~\cite{Culo}. 

The qualitative features of the phase diagram,
similar to those found for the isolated
 system, are as follows.
 For $p > 2$,  the  transition is 
{\it discontinuous} in the sense that the order parameter $q_{\sc ea}$
jumps across the phase boundary. The transition line contains a
{\it tricritical point} $(T^{\star},\Gamma^{\star})$
that divides it in two sections.  For $T \ge T^{\star}$, physical
properties
are {\it continuous} across the transition. The latter is therefore
{\it second order}
in the thermodynamic sense. For $T < T^{\star}$, instead,
  physical quantities
are {\it discontinuous} across the transition which
 is thus {\it first order}. The origin of this behavior is the fact
that the values taken by the parameter $m$ on the transition line are 
 different above and below
$T^{\star}$. For
  $T > T^{\star}$, $m = 1$  along the
transition line. This is its value in the paramagnetic phase meaning
that $m$ is continuous across the
transition and so are the observables.
For $T < T^{\star}$, $m \ne 1$ along the transition
  line but it is a
decreasing function of $T$ that vanishes linearly as $T \to 0$. Crossing
  the phase boundary at $T < T^{\star}$, $m$ is  discontinuous 
  and so are physical properties.
Notice that the line  $T_d(\Gamma,\alpha)$ lies always
{\it above} $T_s(\Gamma,\alpha)$, the static critical line that we 
shall discuss below.

\begin{figure}[t]
\epsfxsize=2.7in
\centerline{
\epsffile{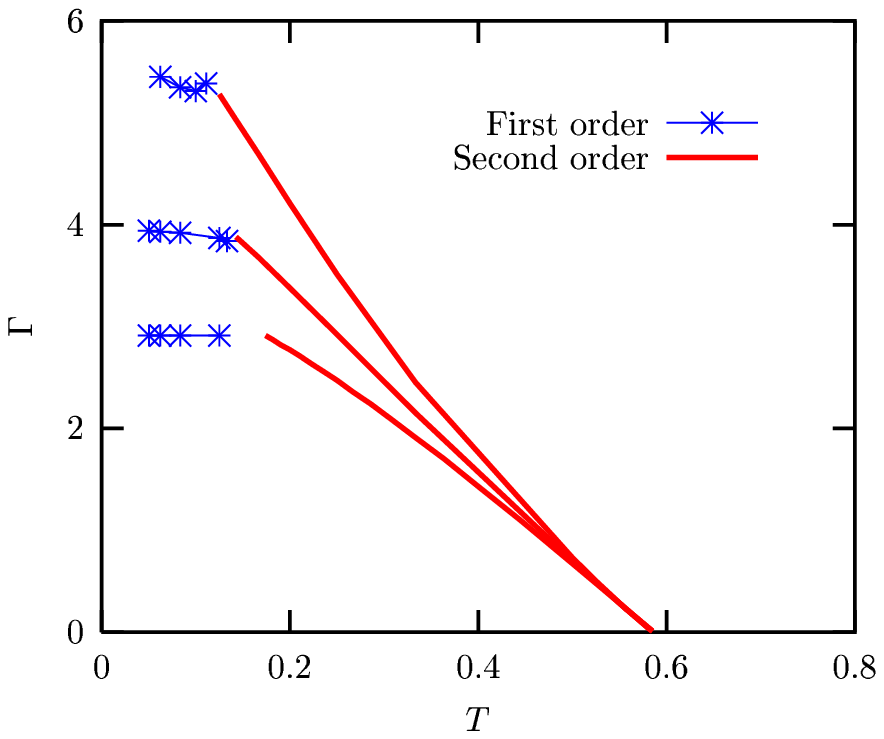}
\hspace{1cm}
\epsfxsize=2.7in
\epsffile{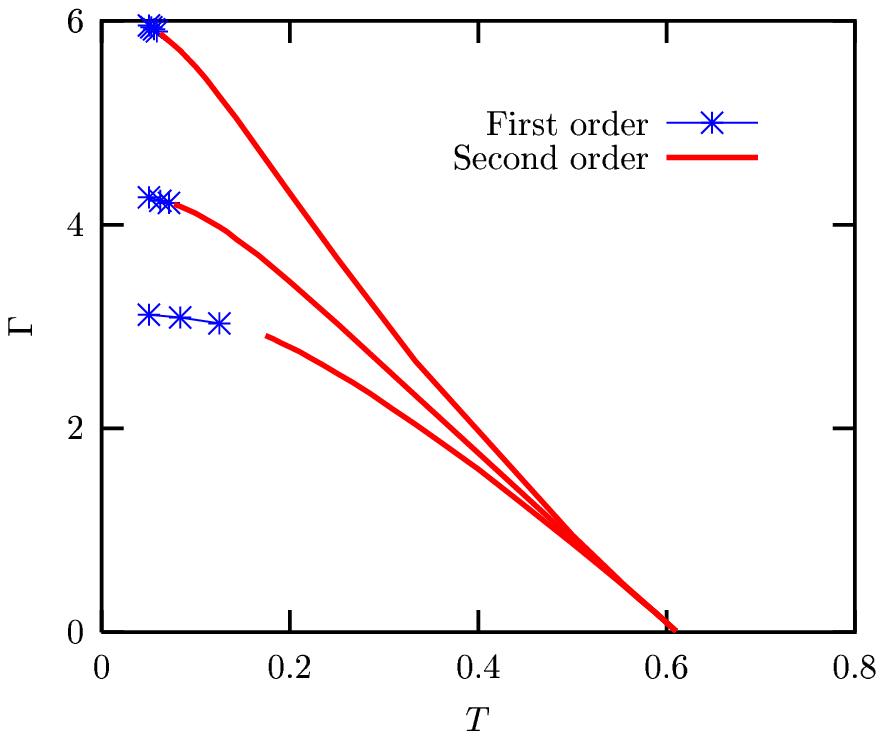}
}
\vspace{.2cm}
\caption{Static (left) and dynamic (right)
phase diagrams for the $p=3$ spin model coupled
to an Ohmic bath ($s=1$). The couplings to the bath are
 $\alpha =$ 0, 0.25, and 0.5 from
bottom to top.
The solid line and line-points represent second and first order transitions,
respectively.}
\label{static}
%\label{dynamic}
\end{figure}

We show on the right panel of  
Fig.~\ref{static} the dynamic phase diagrams
obtained for $p=3$ and three values of the coupling to an Ohmic
bath, $\alpha=0,0.25,0.5$.
The full line and the  line-points represent second and
first order transition, respectively.
We make the following observations:
\begin{enumerate}
\item In the limit $\Gamma \to 0$
 the transition temperature is independent of the strength of the
coupling to the bath.
\item The size of the region in phase space where the system is in the
ordered state increases with $\alpha$.
Coupling to the dissipative environment thus stabilizes this state.
\item
The  dynamic
tricritical temperature decreases rapidly with increasing
$\alpha$.
\end{enumerate}

Our first observation is a consequence of the fact that 
in the limit $\Gamma \to 0$
the partition function is essentially determined
by the zero-frequency components of the pseudo-spin which are
decoupled from the bath (see Section~2). This result is 
however non-trivial from a dynamical point of view, 
since it implies that the dynamic transition of a classical system 
coupled to a colored classical bath is  not modified by the 
latter.

The second feature follows from simple physical considerations.
The interaction term in the action favors spin-glass order.
Coupling to the bath
favors localization and its effect is to reduce the effective
tunneling frequency. 
 Therefore, in the presence of the bath,
the value of the bare tunneling frequency needed to destroy the
ordered state must increase with $\alpha$. 
Even if the localized state and the glassy state 
may seem superficially similar, their are indeed very different.
In the former, the correlation function $C(t+t_w,t_w)$ approaches 
a plateau as a function of $t$ and never decays towards
zero while in the latter the relaxation first approaches a plateau 
but it eventually leaves it to reach zero for $t\gg t_w$. We shall
see this difference explicitly in the analysis of the real-time 
dynamics of Section~4. 

\begin{figure}[t]
\centerline{
\epsfxsize=3in
\epsfysize=2.8in
\epsffile{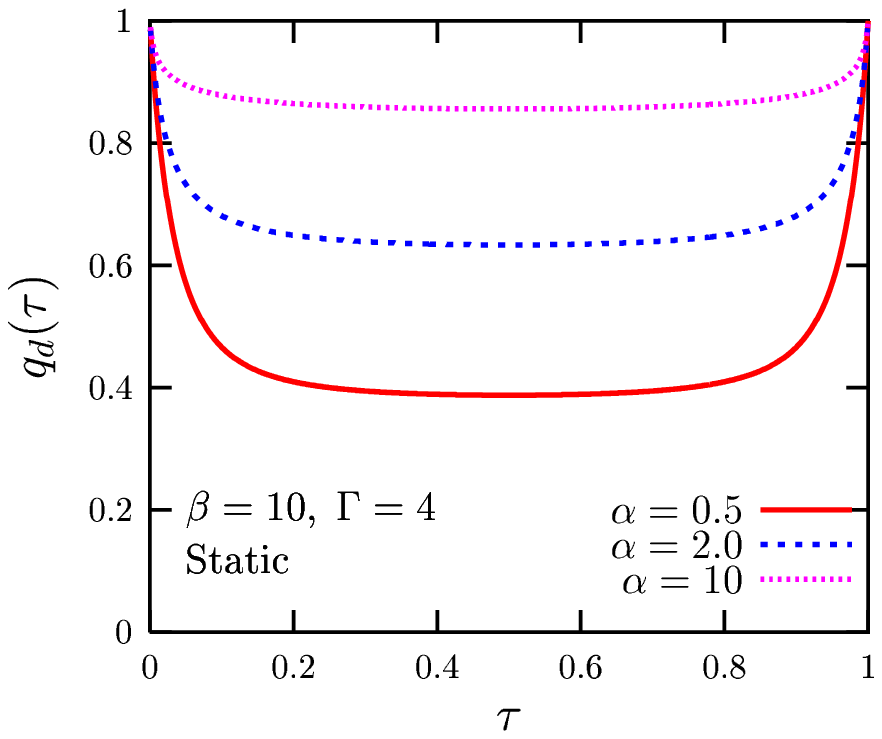}
\hspace{1cm}
\epsfxsize=3in
\epsfysize=2.8in
\epsffile{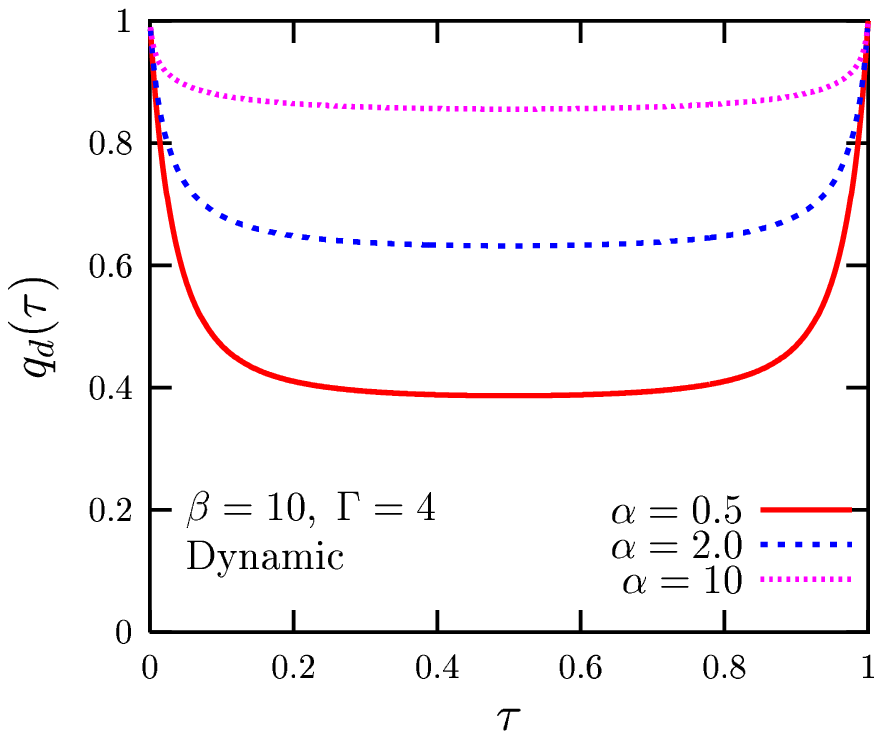}
}
\vspace{.1cm}
\caption{The diagonal part $q_d(\tau)$ for the static (left) and
dynamic (right) solutions}
\label{qd_dynamic}
\end{figure}

\begin{figure}
\epsfxsize=3in
\centerline{\epsffile{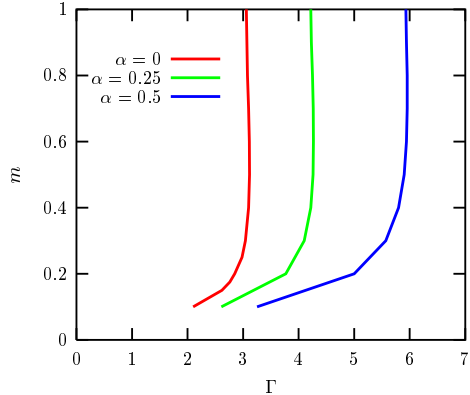}}
\vspace{.2cm}
\caption{The break point $m$ as a function of $\Gamma$ for three
values of the coupling to an Ohmic environment $\alpha$. 
$\beta=20> \beta^\star$.}
\label{m-Gamma}
\end{figure}

The fact that the coupling to the environment favors the ordered state
also reflects itself in the value taken by the order parameters
$q_d(\tau)$ and $q_{\sc ea}$.  
We display in Fig~\ref{qd_dynamic} the $\tau$ dependence of the
diagonal part of the order parameter, $q_d(\tau)$ 
for the static and dynamic
solutions at a fixed temperature and $\Gamma$ 
 for different values of $\alpha$. It can be seen that, 
as $\alpha$ increases, ${q_d}(\tau)$ reaches a higher plateau level at
long imaginary times. The analysis of $q_{\sc ea}$ is postponed to 
Section~3.4 (see Fig.~5).

Figure~\ref{m-Gamma}
displays the $m$-dependence of $\Gamma$ at a fixed temperature 
($T<T^\star$),
for different values of the
coupling to the noise. The function $\Gamma(m)$ is double-valued and
the  physical branch is that on which $dm/d\Gamma > 0$. This is a
consequence of 
Eq.~(\ref{parameters}) that shows that $m$ is a decreasing function of
$q_{\sc ea}$  which itself is a decreasing function of $\Gamma$.
It can be seen that for
 fixed $\Gamma$ and $T$, $m$ decreases with
increasing $\alpha$. Thus, the coupling to the bath results in a
higher effective temperature in the glassy phase (see Section~4
for a definition of $T_{\sc eff}$ and a discussion on this issue). 

\begin{figure}[t]
\epsfxsize=3in \centerline{\epsffile{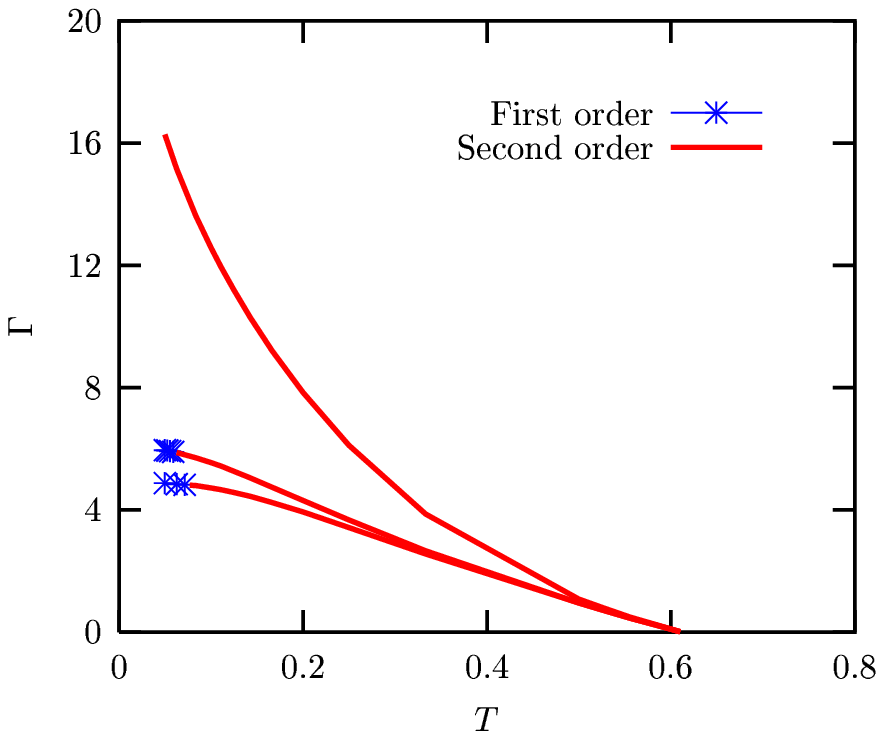}} \vspace{.2cm}
\caption{The dynamic critical line for subOhmic ($s=1/2$, upper curve), 
Ohmic ($s=1$, middle curve) and  superOhmic ($s=1.5$, lower curve) baths. 
$\hbar \omega_{ph}/\tilde{J}=10$ in the non-Ohmic cases. 
The dimensionless coupling to the bath is
$\alpha=0.5$ in all cases. }
\label{Ohmic}
\end{figure}

We have also studied the phase diagram in the non-Ohmic
cases. Figure~\ref{Ohmic} shows a comparison of the effects of an
Ohmic bath and two non-Ohmic ones, subOhmic ($s=1/2$) and superOhmic
($s=3/2$) for the same value of $\alpha$. 
It may be seen that for the chosen values of the parameters 
 the region of stability of the ordered
phase is enhanced (reduced) for a subOhmic (superOhmic) bath
 with respect
to an  Ohmic one. This feature is not generic as there are 
other values of $\omega_{ph}$ for which 
the relative sizes of the effects of Ohmic and nonOhmic baths 
are different. Indeed, in preparing these figures we used 
$\omega_{ph}=10$ in the nonOhmic cases and this parameter 
modifies the coupling to the bath due to the factor $\omega_{ph}^{s-1}$
in $I(\omega)$. 

\subsection{Equilibrium phase diagram.}
 In thermodynamic equilibrium both $q_{\sc EA}$ and $m$
 are determined by imposing that the free energy is an extremum with
respect to their variation. This leads to the conventional
thermodynamic equilibrium state. The value of $x_p$ is obtained from,
\begin{equation}
\label{xpeq}
{x_p^2\over p(1+x_p)} - \log(1+x_p) + {x_p\over (1+x_p)}=0\;.
\end{equation}
 The transition line is defined as
the locus of the
points where the free energies of the {\sc pm} and
{\sc sg} phases coincide.  For each $\Gamma$ and $\alpha$ this defines a
 freezing temperature $T_s(\Gamma,\alpha)$
at which the system enters de {\sc sg} state. The qualitative features of 
the equilibrium phase diagram shown on the left panel of Fig.~\ref{static} 
are  similar to those found for the dynamic case. Notice that
 the line  $T_d(\Gamma,\alpha)$ lies always
{\it above} $T_s(\Gamma,\alpha)$ and that, in contrast to what we found for the dynamic tricritical temperature, the equilibrium tricritical 
temperature $T^{\star}$ depends only 
weakly on the strength of the coupling to the bath.

\subsection{Low-energy properties of the marginal {\sc sg} state}
\label{sec:lowfreq}
\setcounter{equation}{0}
\renewcommand{\theequation}{\thesubsection.\arabic{equation}}
Insights on the low-energy properties of the
model may be gained by studying it in the framework of a simple
and accurate approximation applied to the isolated model in CGS.
It consists in deriving the exact low-frequency form of
$\tilde{q}_{\sc reg}(\omega_k)$ and using it over the whole
frequency range assuming that physical properties at low
temperatures are mainly determined by the low-energy excitations of
the system. We consider in the following the $T=0$ case.

\subsubsection{The low-frequency form of $\tilde{q}_{\sc
reg}(\omega_k)$}
We start by assuming
(and verifying later) that $q_{\sc reg}(\tau)$ (cf.
Eq.~(\ref{regular})) decays in imaginary time as a power-law:
\begin{equation}
\label{powerlaw} q_{\sc reg}(\tau) \propto \left |\tau\right|^{-
\zeta}\;.
\end{equation}
Then, we may write (cf. Eq.~(\ref{sigma})):
\begin{eqnarray}
\label{sigomega}
\tilde \Sigma_{\sc reg}(\omega_k) -
\tilde\Sigma_{\sc reg}(0) &=&
\frac{p}{2}\int_0^{\beta} d\tau \ \left(\cos\omega_k
\tau - 1\right) \left[(p-1) q_{\sc ea}^{p-2} q_{\sc reg}(\tau)
+ \dots \right]\\ \nonumber &\propto& \left| \omega_k \right|^{\zeta-
1}\;\left( 1 + \cdots\right)\;,
\end{eqnarray}
where the dots represent terms that vanish in the limit $\omega_k
\to 0$. Therefore, in the long-time limit,
\begin{equation}
\label{sigmapprox} \tilde \Sigma_{\sc reg}(\omega_k) -
\tilde\Sigma_{\sc reg}(0) \approx \frac{p (p - 1)}{2} q_{\sc
ea}^{p-2} \left[\tilde{q}_{\sc reg}(\omega) - \tilde{q}_{\sc
reg}(0)\right]
\end{equation}
Substituting Eq.~(\ref{sigmapprox}) in Eq.~(\ref{chireg}) and solving
for $\tilde{q}_{\rm reg}(\omega)$ we find:
\begin{equation}
\label{quad} \tilde q_{\sc
reg}(\omega_k)=\frac{2}{\omega_k^2/\Gamma + \alpha_s |\omega_k|^s
+ 2 \kappa_p + \sqrt{\omega_k^2/\Gamma + \alpha_s |\omega_k|^s}
\sqrt{\omega_k^2/\Gamma + \alpha_s |\omega_k|^s+ 4 \kappa_p}},
\end{equation}
where we introduced the parameter
\begin{equation}
\label{kappa}
\kappa_p \equiv \sqrt{\frac{p(p-1) q_{\sc ea}^{p-2}}{2}}
\; .
\label{kappa_def}
\end{equation}
Equation~(\ref{quad}) only holds in the low-frequency limit where it
reduces to
\begin{equation}
\label{lfl} \tilde q_{\sc reg}(\omega_k) = \kappa_p^{-1}\;\left[ 1
- {1 \over \sqrt{\kappa_p}}\; \left(\omega_k^2/\Gamma + \alpha_s
\left| \omega_k \right|^s \right)^{1/2}\;\right]\;\;\;\to
\kappa_p^{-1}\;\left[ 1 - \sqrt{{\alpha_s \over\kappa_p}}\;
 \left|\omega_k\right|^{s/2}\right]\;.
\end{equation}
This leads to the long-$\tau$ behavior
\begin{equation}
\label{ltl} q_{\sc reg}(\tau) \sim \sqrt{{\alpha_s
\over\kappa_p^3}}\; {1 \over \left| \tau \right|^{1 + {s/2} }}\;.
\end{equation}
The assumption (\ref{powerlaw}) is thus self-consistent with the
exponent $\zeta = 1 + s/2$. In the absence of the bath
Eq.~(\ref{quad}) leads to
\begin{equation}
\label{ltlnb} q_{\sc reg}(\tau) \sim {1 \over \sqrt{\Gamma
\kappa_p^3}}\; {1 \over \left| \tau \right|^2}\;,
\end{equation}
the result found previously in CGS for the isolated system. A
crossover between these two regimes occurs at $\tau_{cr} =
\left(\Gamma\alpha_s\right)^{1/(s-2)}$. The analytic continuation
of Eq.~(\ref{lfl}) yields the low-frequency limit of the imaginary
part of the susceptibility:
\begin{equation}
\label{imchilfl} \chi_{\sc reg}''(\omega) \sim {\rm sign}(\omega)
\left( {\alpha_s \over \kappa_p^3} \right)^{1/2} \left| \omega \right| ^{s/2}.
\end{equation}
The result for the Ohmic case, $\chi_{\sc reg}''(\omega) \propto
\left| \omega \right| ^{1/2}$ was previously found for $p=2$
continuous and discrete Kondo-alloy models at the quantum critical 
point~\cite{metallic}. In the
marginally stable state, this behavior persists throughout the
low-temperature phase.

Equation~(\ref{quad}), exact in the limit $\omega_k \to 0$, may be used as
an approximation for finite frequencies.  It will be seen in the
following that this approximation allows one to gain useful insight on
the effects of the environment on the physics of the interacting system.

\subsubsection{The quantum phase transition}
\label{qpt}
The normalization condition and Eq.~(\ref{regular}) lead to the
following
equation for the order parameter $q_{\sc ea}$ at $T=0$:
\begin{equation}
\label{norm}
1-q_{\sc ea} =\frac{1}{\beta}
\sum_{\omega_k} \tilde q_{\sc reg}(\omega_k)
\; \mathop{\rightarrow} \limits_{T = 0}
\int_{-\infty}^\infty\;{d\omega \over 2 \pi}\;\tilde q_{\sc reg}(\omega)
\; ,
\label{qea_eq}
\end{equation}
This is still an implicit
equation for the order parameter
as $\tilde q_{\sc reg}(\omega)$ depends on  $q_{\sc
ea}$ through $\kappa_p$ (cf. Eq.~(\ref{kappa_def})).

We now approximate Eq.~(\ref{qea_eq}) by assuming that the
integral on the right-hand side is dominated by the low
frequencies. Then, we use for $q_{\sc reg}(\omega)$ the expression
given in Eq.~(\ref{quad}) which is asymptotically exact in this
limit and write:
\begin{eqnarray}
1-q_{\sc ea} = \frac{2}{\pi}
\int_0^\infty \;
\frac{d\omega}{\omega^2/\Gamma + \alpha_s \omega^s + 2 \kappa_{p}
+\sqrt{\omega^2/\Gamma + \alpha_s \omega^s}
\sqrt{\omega^2/\Gamma + \alpha_s \omega^s + 4 \kappa_{p} }  }
\; .
\label{int}
\end{eqnarray}
It is convenient to make the change of variables
$\omega = (\Gamma \alpha_s)^{1/(2 - s)} x$  in Eq.~(\ref{int}) which leads
to:
\begin{equation}
\label{int_sub}
A_s\;(1 - q_{\sc
ea}) = \int_0^\infty \;
\frac{dx}{x^2 + x^s + 2 \epsilon
+\sqrt{x^2 + x^s}
\sqrt{x^2 + x^s + 4 \epsilon }  } \equiv g_s(\epsilon)\;,
\end{equation}
where
\begin{eqnarray}
\label{Aandeps_def}
A_s&=&{\pi \over 2}\;\left(\Gamma^{s-1} \alpha_s\right)^{1\over 2 -
s}\; = {\pi \over 2} \left({\hbar \over M \omega_{ph} }\right)^{s -
1\over 2 - s}\;\left(\alpha \over \sin \pi s/2 \right)^{1\over 2- s }\;,\\
\nonumber\\\epsilon&=& {\kappa_p \over \left(\Gamma^s \alpha_s^2\right)^{1\over 2
- s}} = {\tilde{J} \over \hbar \omega_{ph}}
 \left[
 \left({\hbar \over M \omega_{ph}} \right)^s
 {\alpha^2\over \sin^2 \pi s/2}
\right]^{1 \over s - 2}\;\kappa_p
\; .
\end{eqnarray}
Equation~(\ref{int_sub}) will be used to study the $T=0$ quantum phase
transition. We shall mostly be interested in the 
vicinity of the quantum transition where the system is close
to the quantum paramagnetic state. We discuss separately different
types of environment.

\vspace{.5cm}
\noindent{\it The Ohmic case}
\vspace{.5cm}

Setting $s=1$ in Eq.~(\ref{int_sub}) the equation of state may be written as:
\begin{equation}
\label{int_ohmic} {\pi \alpha \over 2}  (1-q_{\sc ea}) =
\int_0^\infty \; \frac{dx}{x^2 + x + 2 \epsilon +\sqrt{x^2 + x}
\sqrt{x^2 + x + 4 \epsilon }  } \; \equiv g_1(\epsilon),
\end{equation}
with $\epsilon = \kappa_p/\left(\Gamma \alpha^2\right)$.

\begin{figure}[t]
\epsfxsize=5in \centerline{\epsffile{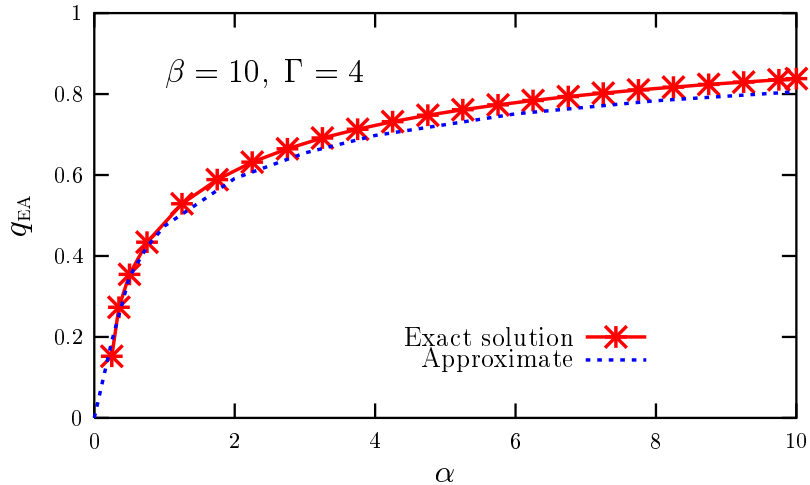}}
\caption{The dynamic Edwards-Anderson parameter as a
function of the coupling to an Ohmic bath, $\alpha$, for
$\Gamma=4$ and $\beta=10$.  Solid line with point: exact numerical
calculations. Dashed line: the low-frequency approximation of
Eq.~(\protect{\ref{int}}).}
\label{qeag}
\end{figure}

We show in Fig.~\ref{qeag} the $\alpha$-dependence of $q_{\sc ea}$
 in the marginally stable case for $p=3$ at fixed $\Gamma = 4$.
We represent with line-points the results obtained numerically from
the full
 equations at a
 finite but low temperature, $T=0.1$. The dashed line instead
represents the approximate solution derived from 
Eq.~(\ref{int}). The agreement between the two
calculations is very good even if the approximation strictly applies 
to the  zero temperature case only.

\begin{figure}
\epsfxsize=3in \centerline{\epsffile{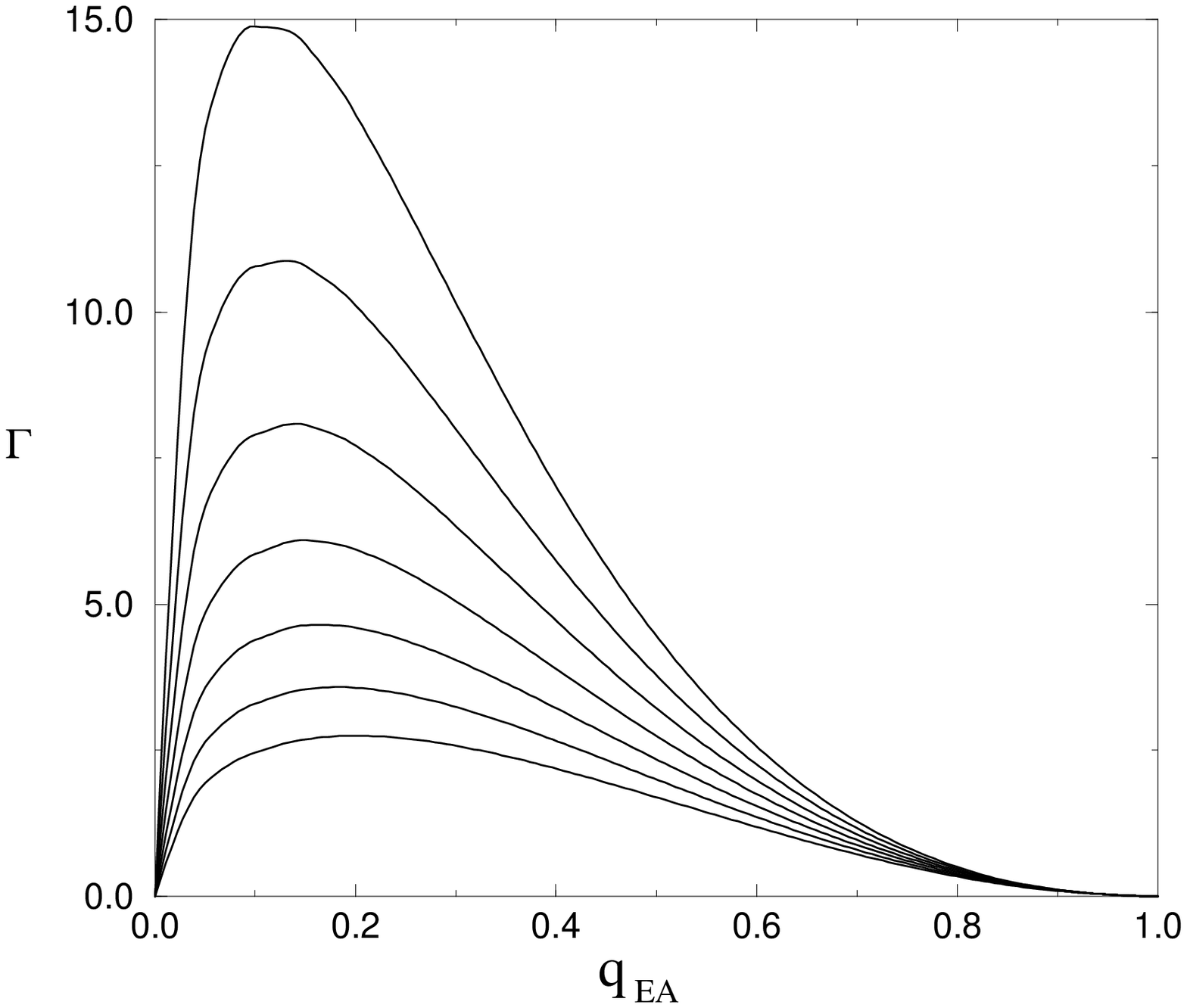}}
\caption{$\Gamma$ as a function of $q_{\sc ea}$ at  $T=0$ for
$p=3$ and an Ohmic bath. The curves follow from the numerical 
solution to Eq.~(\ref{int_ohmic}). The coupling
to the bath $\alpha$ runs from $0$ to $1.2$ in intervals of $0.2$
from bottom to top.} \label{gam_qea}
\end{figure}

Figure~\ref{gam_qea} shows $\Gamma$ as a function of $q_{\sc ea}$
for $p=3$ and several values of $\alpha$ as obtained from the numerical
solution to Eq.~(\ref{int_ohmic}). The $T=0$ transition takes
place at the maximum value of $\Gamma$, $\Gamma_d$. The
 corresponding value of $q_{\sc ea}$
is the discontinuity of the order parameter at the first-order
transition. While $\Gamma_d$ increases rapidly with $\alpha$, the
jump of the order parameter {\it decreases} as the strength of the
coupling to the bath increases. The presence of the Ohmic bath
thus tends to make the first order transition smoother.
(This property tells us that it will be very difficult to
see the first order transition by solving numerically the real-time 
dynamic equations.)

This behavior results from the fact that $g_1(\epsilon)$
 diverges logarithmically
as $\epsilon \to 0$. In order to see this, we decompose the
interval of integration in two parts, $0 \le x \le 1$ and $1 \le x
\le \infty$. The integral over the second interval is a finite
constant at $\epsilon=0$. In the integral over the first interval
we may neglect $x^2$ compared to $x$ and write
\begin{equation}
\label{int_ohmic_approx}
{\pi \alpha \over 2}(1-q_{\sc ea}) \mathop{\sim} 
\limits_{\epsilon \to 0}\;\int_0^1\;
\frac{dx}{ x + 2 \epsilon
+\sqrt{ x}
\sqrt{ x + 4 \epsilon }  }
 \; = 
- {1 \over 2} \ln \epsilon + {\cal O}\left( 1\right).
\end{equation}
 We choose $p=3$ for concreteness  and solve
Eq.~(\ref{int_ohmic_approx}) for $\Gamma$ to obtain the equation
of state in the high noise limit:
\begin{equation}
\label{eqofstate} \Gamma = {1 \over \alpha^2} \sqrt{3\;q_{\sc
ea}}\;e^{\pi \alpha (1 - q_{\sc ea})}\;.
\end{equation}
This function has a maximum at
\begin{equation}
\label{qeamax} q_{\sc ea}^{\star} \approx {1 \over 2 \pi \alpha}\;,
\end{equation}
where $\Gamma$ reaches the value
\begin{equation}
\label{Gammamax} \Gamma_{\rm max} \equiv \Gamma_d \approx {\sqrt{3
\over 2 \pi \alpha^5}} \exp \pi \alpha\;.
\end{equation}
We thus find the two features mentioned above, namely, a reduction
of the discontinuity of  the order parameter and a rapid
increase of $\Gamma_d$ for high values of $\alpha$. Expressing
Eq.~(\ref{Gammamax}) in terms of the original variables
  of the problem (cf. Eq.~(\ref{units})) we find that, in the high
  noise limit, the $T=0$ dynamic freezing
transition takes place at the critical coupling
\begin{equation}
\label{critical_coupling} \tilde{J}_d \sim {\hbar \alpha^{5/2}
\over M} \exp( -\pi \alpha) \;.
\end{equation}
Thus, for $\alpha \gg 1$, $\tilde{J}_d$ is proportional to the
exponentially small energy scale of
Eq.~(\ref{tls_limit}) associated to incoherent tunneling in the
isolated {\sc tls}.
It must be emphasized that the existence of this
scale is a feature of the spherical model used in this paper.  
Real {\sc tls} ({\it i.e.}, described by Ising spins) localize
at $\alpha = 1$. Therefore, $\tilde{J}_d$ is expected to vanish {\it
precisely} at  $\alpha = 1$ for discrete spins.   

Deep in the ordered phase the system is expected to freeze with  $q_{\sc ea}
\approx 1$. This regime occurs for sufficiently high values of
$\alpha$ or sufficiently low values of $\Gamma$.  
Consider first the former case with $\Gamma \alpha^2 \gg 1$. Then,
 $\epsilon \ll 1$ and we can still 
use Eq.~(\ref{int_ohmic_approx}) which, for  $q_{\sc ea} \approx 1$, reduces 
to 
\begin{equation}
\label{qeanear1ohmic}
q_{\sc ea} \approx 1- {1 \over \pi \alpha} \ln \left({\Gamma \alpha^2
\over \kappa_p(1)}\right)\;,\;\;\;\Gamma \alpha^2 \gg 1\;.
\end{equation}
where $ \kappa_p(1) = \sqrt{p (p-1)/2}$.

In the opposite case, $\Gamma \alpha^2 \ll 1$, $\epsilon$ is large. 
In this case
\begin{equation}
g_1(\epsilon) = {2 \over 3 \sqrt{\epsilon}}\;\left( 1 - {3 \over 8
\sqrt{\epsilon}} + \cdots \right)\;,
\end{equation}
leading to
\begin{equation}
\label{qeanear1ohmic2}
q_{\sc ea} \approx 1- {4 \over 3 \pi}\;\left({\Gamma \over
\kappa_p(1)}\right)^{1/2}\;\left[ 1 - {3 \over 8}\left({\Gamma \over
\kappa_p(1)}\right)^{1/2} \alpha + \cdots \right]\;,\;\;\;\Gamma
\alpha^2 \ll 1\;.
\end{equation}

In both regimes the effect of the noise leads to an increase in $q_{\sc ea}$
thus stabilizing the ordered phase.

\vspace{.5cm}
\noindent{\it The subOhmic case}
\vspace{.5cm}

\begin{figure}
\epsfxsize=3in
\centerline{\epsffile{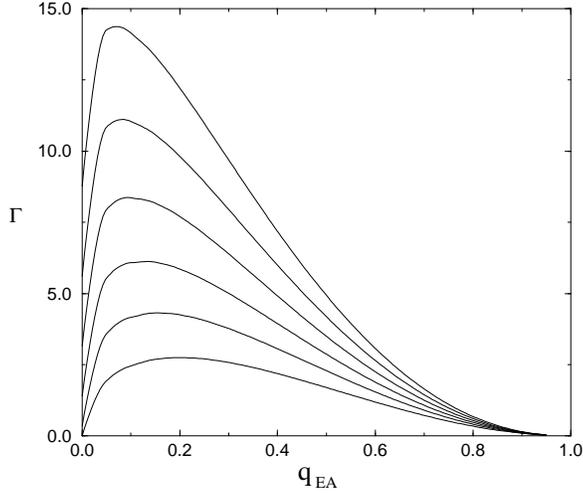}}
\caption{$\Gamma$ as a function of $q_{\sc ea}$ at $T=0$ for $p=3$ and
a subOhmic bath, $s=1/2$
from Eq.~(\ref{int}). The coupling $\alpha_{1/2}$
runs from $0$ to $2$ in intervals of $0.4$ from bottom to top.}
\label{gam_qea_sub}
\end{figure}

Figure~\ref{gam_qea_sub} shows $\Gamma$ as a function of $q_{\sc ea}$ for
$p=3$ and
several values of the coupling to a subOhmic bath with $s=1/2$. The
results were obtained by numerically
solving Eq.~(\ref{int}). The qualitative features of these curves
are similar to
those found in the Ohmic case.

As discussed in Section \ref{themodel}, in the subOhmic case the
isolated {\sc tls} has a localization transition at a
critical value $\alpha^{\sc crit}$ of the
coupling to the bath.  We thus expect a transition to the 
ordered phase at $\tilde{J} = 0$ for all $\alpha > \alpha^{\sc
crit}$  in the interacting system.
Near the critical point at $\tilde{J} = 0$, $\epsilon$ is
small. For $s < 1$ the integral on the right-hand side of Eq.~(\ref{int_sub})
is finite at $\epsilon = 0$ and $g_s(0) = f_s(0)/2$
where $f_s$ is the function defined in Eq.~(\ref{const_tls}).
Detailed inspection of the behavior of $g_s(\epsilon)$ shows that, as $\epsilon
\to 0$
\begin{eqnarray}
\label{limit_g}
 g_s(0) -  g_s(\epsilon) \propto \left\{
\begin{array}{llc}
\epsilon & {\rm for} & 0 < s < 1/2 \\
\epsilon \ln\left( 1/\epsilon\right)  & {\rm for} &  s = 1/2\\
\epsilon^{{1 - s \over s} } & {\rm for} & 1/2 <  s < 1
\end{array}
\right.
\end{eqnarray} 

We consider for simplicity the case $s < 1/2$ and take
$p=3$  for concreteness. For $\alpha \sim \alpha^{\rm crit}$,  
Eq.~(\ref{int_sub}) acquires the form
\begin{equation}
\label{}
{\tilde{J} M \over \hbar^2} \sqrt{q_{\sc ea}} \propto 1 - \left({\alpha \over
\alpha^{\rm crit}}\right)^{1 \over 2-s}\;(1 - q_{\sc ea}),
\end{equation}
where $\alpha^{\rm crit}$ is given in Eq.~(\ref{loc_tra}).
There is a maximum at $q_{\sc ea}
 \propto \left(\tilde{J} M /\hbar^2\right)^2$. The dynamic
 transition thus takes place at
\begin{equation}
\label{jcrit_sub}
\tilde{J}_d  \propto  {\hbar^2 \over M}\;\left(1 - {\alpha \over
\alpha^{\rm crit}}\right)^{1/2}\;.
\end{equation}
The jump of the order parameter at the transition is $q_{\sc ea}^{\star}
\propto ( 1 - \alpha / \alpha^{\rm crit} )$. Therefore, for $\alpha = \alpha^{\rm crit}$ 
the dynamic transition is {\it continuous}. 
Summarizing, for $\alpha<\alpha^{\sc crit}$ the transition between
{\sc pm} phase and {\sc sg} phase occurs at a finite value of $J_d$ 
while for $\alpha > \alpha^{\sc crit}$ an infinitesimal $\tilde J$ is
enough to render the system glassy. We expect to obtain this same
behavior for an interacting {\sc tls} in an Ohmic bath.

At large couplings,  $\alpha \gg \alpha^{\rm crit}$, $q_{\sc ea} \sim 1$ and we find
\begin{equation}
\label{qeanear1nonohmic}
q_{\sc ea} \approx 1- \left( {\alpha^{\rm crit} \over \alpha}
\right)^{1\over 2-s}\;.
\end{equation}
Notice the absence of $\tilde{J}$-dependent corrections that
 appear
 at higher order ($\alpha^{-2}$).

In the opposite limit (large $\tilde{J}$), $\epsilon$ is large. Then, 
\begin{equation}
\label{gslargeeps}
g_s(\epsilon) = 
{2 \over 3 \sqrt{\epsilon}}\;\left( 1 - a_s
\epsilon^{(s/2 -1)} + \cdots \right)\;,
\end{equation}
with $a_s$ a constant. We find
\begin{equation}
\label{nonohmic_smallnoise}
 q_{\sc ea} \approx 1- {4 \over 3 \pi}\;\left({\Gamma \over
\kappa_p(1)}\right)^{1/2}\;\left[ 1 - a_s {\Gamma^{s/2} \over
\kappa_p^{1 - {s\over 2}}(1)}\; \left({ \hbar
\omega_{ph}  \over \tilde{J}}\right)^{1 - s}
\;{\alpha \over \sin {\pi s \over 2}} + \cdots 
\right] 
\end{equation}
As before, the presence of noise favors the ordered phase. The 
comparison
of Eqs.~(\ref{qeanear1ohmic}) and (\ref{qeanear1nonohmic}) shows that,
at strong coupling, an Ohmic bath is more effective than a subOhmic
bath in freezing the spins. At weak coupling we have a linear
dependence on $\alpha$ in both cases. For $\Gamma \ll 1$, however, the slope is
higher in the latter case which results now in higher values of $q_{\sc
ea}$. Notice the presence of the extra factor  $\left(\hbar
\omega_{ph}/\tilde{J}\right)^{1-s}$ that amplifies this effect if the
phonon energy is larger than the magnetic energy.

\vspace{.5cm}
\noindent{\it The superOhmic case}
\vspace{.5cm}

In the superOhmic case no localization transition exists at $\tilde{J} = 0$. 
For $s > 1$, $g_s(\epsilon)$ diverges as $\epsilon^{{1 - s \over s}}$
in the limit $\epsilon \to 0$. This corresponds to small $\tilde{J}$ 
or large $\alpha_s$. A calculation similar to those performed above
yields the critical coupling 
\begin{equation}
\label{eq_of_state_low_J} 
\tilde{J}_d \sim \hbar \omega_{ph} \left( {\sin {\pi s \over 2}\over
\alpha}\right)^{s - 1}\;,\;\;\;\alpha \gg 1\;.
\end{equation}
As in the Ohmic case, the critical coupling decreases with increasing
$\alpha$ but only as a power law. The jump of the order parameter 
at the transition is however independent of $\alpha$. 

Deep in the ordered phase, for small values of $\tilde{J}$, we find
\begin{equation}
\label{low_J_large_noise} 
q_{\sc ea} \sim 1 - \left({\tilde{J}_d \over \hbar \omega_{ph}}\right)^{{1 - s
\over s}} \left( {\sin {\pi s \over 2}\over
\alpha}\right)^{s - 1}\;.
\end{equation}

For small values of $\alpha$ Eq.~(\ref{nonohmic_smallnoise}) is still
valid. Notice that for $s >1$ the enhancement of the order parameter
due to the coupling to the
bath decreases when $\omega_{ph}/\tilde{J}$ increases.

\subsection{The real-time correlation function}
\label{real-time}
\setcounter{equation}{0}
\renewcommand{\theequation}{\thesubsection.\arabic{equation}}

In thermodynamic equilibrium the correlation function and the
imaginary part of the susceptibility are related by
\begin{equation}
\label{corr_gen}
C(t) \equiv {1 \over N} \sum_i \langle s_i(t) s_i(0) \rangle = q_{\sc ea} + \hbar \int_0^{\infty} \frac{d\omega}{\pi}
\chi_{\sc reg}''(\omega)
\coth(\beta \hbar \omega/2) \cos(\omega t).
\end{equation}

If instead of the equilibrium response function we use in Eq.~(\ref{corr_gen})
the expression for $\chi''(\omega)$ obtained through the {\sc ams} we obtain
a correlation function that is closely related to the {\it stationary}
part of that obtained through real-time dynamical calculations. This
relationship was discussed extensively in CGS in the case of the
isolated system with the following conclusions:
\begin{enumerate}

\item  The temperature $T_d$ below which the {\sc ams}
solution exists coincides precisely with
the dynamical critical temperature obtained from the dynamical
calculations. This is the temperature below which 
the real-time dynamics of the system becomes non-stationary and
violations of
FDT  set in. 

\item
The parameter $m$ precisely coincides with $X$,
the FDT violation factor. This is
related to the {\it effective} temperature of the system
in the aging regime, $T_{\rm eff} = T/X$~\cite{Cukupe}, see 
Section~4.

\item The response function derived from the {\sc ams} is identical to
 the out-of-equilibrium response function 
when the long waiting-time is taken first and the weak-coupling
limit in taken later on. More precisely,
\begin{equation}
C_{\sc ams}(t) \equiv {1 \over N} \sum_i \langle s_i(t) s_i(0)
\rangle_{\sc ams} =  \lim_{\alpha \to 0} \lim_{t_w \to
\infty} C_{\sc dyn}(t + t_w,t_w)\;.
\label{relationship}
\end{equation}
\end{enumerate}

A proof of the analogous properties for the system coupled to the bath 
can be given  following the same
lines. The first two conclusions remain unchanged. The third one
generalizes to 
\begin{equation}
C_{\sc ams}(t) =  \lim_{t_w \to
\infty} C_{\sc dyn}(t + t_w,t_w)\;.
\label{identity}
\end{equation}
valid for all values of $\alpha$. The aging regime, $t \gg t_w$, in which $C_{\sc
dyn} (t + t_w,t_w)$ decreases below $q_{\sc ea}$ 
is not accessible in this approach.

In this Section we shall analyze in detail several time regimes 
in $C(t)$. We use throughout
this section the original variables of the problem.

\subsubsection{No coupling to the bath}
\label{nocoupling}
We consider first the case in which there is no coupling to a bath. Then, the analytic continuation of
Eq.~(\ref{quad}) is
\begin{equation}
\label{xi2nonoise}
\chi''_{\sc reg}(\omega) = {\omega \over 2 \kappa_p^2} \left( {M \over
\tilde{J}^3} \right)^{1/2} \sqrt{4 \kappa_p - {\omega^2 M \over
\tilde{J}}}\;.
\end{equation}
Substituting this expression in Eq.~(\ref{corr_gen}) and making the
change of variables $\omega = \sqrt{4 \kappa_p \tilde{J}/M}\ x$ in the
integrals we obtain the correlation function
\begin{equation}
\label{corr_zero_noise}
C(t)=q_{\sc ea}+ \frac{2 \hbar}{\pi \kappa_p} \left(\frac{4
\kappa_p}{M \tilde{J}}\right)^{1/2} \int_0^1 dx\ x\ \sqrt{1 - x^2} \cos(x
t/t_0)  \coth\left(\frac{\hbar}{2 T t_0}x\right),
\end{equation}
where $t_0$ is a characteristic time given by
\begin{eqnarray}
\label{time}
t_0&=&\left(\frac{M}{4 \kappa_p \tilde{J}}\right)^{1/2}
\; .
\end{eqnarray}

At $T=0$  Eq.~(\ref{corr_zero_noise})
reduces to
\begin{eqnarray}
\nonumber C(t) &=&
q_{\sc ea}+  \frac{2 \hbar}{\pi \kappa_p} \left(\frac{4
\kappa_p}{M \tilde{J}}\right)^{1/2} \int_0^1 dx\ x\ \sqrt{1 - x^2} \cos(x
t/t_0)\\
\label{corr_T=0}\\
\nonumber &=& q_{\sc ea}+  \frac{2 \hbar}{3 \pi \kappa_p} \left(\frac{4
\kappa_p}{M \tilde{J}}\right)^{1/2}\   _1F_2(1;1/2,5/2;-(t/t_0)^2/4)
\; ,
\end{eqnarray}
where $ _1F_2$ is a generalized hypergeometric function.
From the normalization condition $C(t) = 1$ we
find the quantum equation of state,
\begin{equation}
\label{qEA_T=0}
\kappa_p (1-q_{\rm EA}) = \frac{2\hbar}{3 \pi} \left(\frac{4 \kappa_p}{M \tilde{J}}\right)^{1/2},
\end{equation}
found previously in CGS.
The asymptotic behavior of the correlation function in the long-time
limit is
\begin{equation}
\label{corr_as_0}
C(t) \mathop{\rightarrow} \limits_{t \gg t_0}  q_{\sc ea} +
 \frac{2 \hbar}{3 \pi \kappa_p} \left(\frac{4
\kappa_p}{M \tilde{J}}\right)^{1/2} \left(\frac{t_0}{t}\right)^{3/2} f(t/t_0)
\; ,
\end{equation}
where $f$ is an oscillatory  function that can be expressed in
terms of Fresnel integrals. From Eqs.~(\ref{qEA_T=0}) and
(\ref{time}) the frequency of the oscillations is
\begin{equation}
\label{w0}
\omega_0 \sim {\hbar \over M (1 - q_{\sc ea})}\;.
\end{equation}
At the dynamic transition point $q_{\sc ea}$ depends only on $p$.
Then, $\omega_0 \propto \hbar/M$, the characteristic frequency of
the non-interacting {\sc tls}. Deep in the ordered phase $q_{\sc ea}
\approx 1$ and Eq.~(\ref{qEA_T=0}) yields $1- q_{\sc ea} \sim
\hbar/\sqrt{M \tilde{J}}$. Then, in this limit $\omega_0 \sim
\sqrt{\tilde{J}/M}$ .

At temperatures higher than $ T_{cr}= \hbar/t_0 ~\sim \hbar
\sqrt{\tilde{J}/M}$ , but low so that the results from the 
approximation can still be used, we can approximate
$\coth z \sim z^{-1}$ in the integral on the right-hand side of
Eq.~(\ref{corr_zero_noise}) and write
\begin{eqnarray}\nonumber
C(t) &\sim& q_{\sc ea}+ \frac{4 T}{\pi \tilde{J} \kappa_p}
\int_0^1 dx\ \sqrt{1 - x^2} \cos\left(\frac{x t}{t_0}\right)\\
\label{corr_Tneq0}\\
\nonumber &=&
q_{\sc ea}+ \frac{2 T}{\tilde{J} \kappa_p} J_1\left(\frac{t}{t_0} \right) \frac{t_0}{t}
\; ,
\end{eqnarray}
where $J_1(x)$ is the Bessel function. Notice that
Eq.~(\ref{corr_Tneq0}) also holds for {\it all} temperatures for
times $t \gg \hbar/T$.
The  normalization condition now yields
\begin{equation}
\label{qEA_heavy}
\kappa_p (1-q_{\rm EA}) =  \frac{4 T}{\pi \tilde{J}} \int_0^1 dx\ \sqrt{1 -
x^2}={T \over \tilde{J}} \; ,
\end{equation}
which is the {\it classical} equation for $q_{\rm EA}$~\cite{Cuku}. 
In this classical regime the long-time asymptotic behavior of the
correlation function is
\begin{equation}
\label{corr_heavy_ass}
C(t) \mathop{ \rightarrow } \limits_{t \gg t_0} q_{\sc ea}
+  \frac{2 T}{\tilde{J} \kappa_p} \sqrt{\frac{2}{\pi}} \left(\frac{t_0}{t}
\right)^{3/2} \cos\left(\frac{3\pi}{4} - \frac{t}{t_0}\right).
\end{equation}
Notice that the power-law decay of the amplitude of the oscillations
$\propto t^{-3/2}$  at high and low temperatures is the same.

\subsubsection{Finite coupling to a bath.}
\label{non_zero_noise}

In the presence of a coupling to an Ohmic bath there are two
different regimes. At frequencies higher than $\omega^{\star} =
\hbar \alpha/M$ the inertial term in Eq.~(\ref{quad}) dominates
over the term proportional to $\alpha$. For times shorter than
$t^{\star} = 1/\omega^{\star}$ the system thus behaves as if it
were isolated. At longer times, when inertia may be ignored, we
have
\begin{equation}
\label{chi"_no_inertia} \chi''(\omega) \sim {1 \over \hbar}
{\sqrt{\alpha \omega} \over \alpha \omega + 2 \kappa_p
\tilde{J}/\hbar}\; \left(\;{\hbar \over 2 \kappa_p
\tilde{J}}\;\right)^{1/2}\;.
\end{equation}
and the motion is overdamped. The correlation function then reads
\begin{equation}
\label{C_no_inertia} C(t) \mathop{\rightarrow} \limits_{t \gg 
t^{\star}}\;q_{\sc ea} + {2 \hbar \sqrt{\gamma_0} \over \pi
\;\left(\; 2 \kappa_p \tilde{J}\;\right)^{3/2}} \int_0^\infty
d\omega\;\sqrt{\omega} \cos \omega t \coth\left(\beta \hbar
\omega/2\right)\;,
\end{equation}
where $\gamma_0 =\alpha \hbar$ is the classical friction
coefficient. Performing the integral we find
\begin{eqnarray}
\label{c_no_inertia_asymptotic} C(t) - q_{\sc
ea}\;\mathop{\rightarrow} \limits_{t \gg t^{\star}}\;{2 \hbar
\sqrt{\gamma_0/\pi} \over \left(\; 4 \kappa_p
\tilde{J}\;\right)^{3/2}} \times \;\left\{
\begin{array}{ll}
- t^{- {3\over 2}} \;, & T = 0\;,\\
\\
 4 T/\hbar\; t^{- {1 \over 2}} \;, & T \gg \hbar/t\;.
\end{array}
\right.
\end{eqnarray}
Notice the difference in  sign between the results at zero and
finite temperature. At zero temperature $C(T)$ approaches $q_{\sc ea}$ from
below, whereas at $T\ne 0$ it does so from above. In the Ohmic case 
the exponent
controlling the decay of the $T=0$ correlation function is the
same that controls the amplitude of the coherent oscillations
found in the absence of noise.

At finite temperature the decay is slower, $C(t) - q_{\sc ea}
\propto t^{- 1/2}$ . In the classical model, the non-equilibrium
correlation function  $C(t + t_w, t_w)$  approaches the plateau
$q_{\sc ea}$ as $C(t + t_w, t_w) - q_{\sc ea} \propto t^{-\nu(T)}$
for $t \ll t_w$. It was found~\cite{Culedou} that the
temperature-dependent exponent $\nu(T)$ approaches $1/2$ 
in the zero temperature limit in agreement with our result. The calculation of
the temperature corrections to the exponent lies beyond the power
of our low-temperature approximation.

At finite temperature, in the long-time limit,  our
results coincide  with those obtained
from the solution of the classical Langevin equation without
inertia. Although the asymptotic form of the correlation function
is independent of  $M$ ({\it i.e.}, of the tunneling frequency
$\Delta$) it must be remembered that
  Eq.~(\ref{c_no_inertia_asymptotic})
 only holds for times longer than $t^{\star}$ which does depend 
on $\Delta$. A consequence of this fact is that the dynamics 
 of the model  in the limit 
$\Delta \to 0$ is trivial. Indeed, it can be shown 
from Eq.~(\ref{quad}) that
\begin{equation}
\label{classical_limit} \chi''_{\sc reg}(\omega) \mathop{\to}
\limits_{M\omega/\gamma_0 \gg 1} {\gamma_0 \over M^2
\omega^3}\;.
\end{equation}
Then, for any finite $\omega$,
\begin{equation}
\label{imchi3} \lim_{M/\gamma_0 \to \infty}\chi_{\sc
reg}''(\omega)\equiv 0 \; .
\end{equation}
However $\chi_{\sc reg}''(\omega)$ cannot be identically zero
since the static susceptibility, $\chi_{\sc reg}(0)$, 
is finite and it is given by $\chi_{\sc reg}(0)=\kappa_p^{-1}$
according to Eq.\ (\ref{quad}). 
$\chi_{\sc reg}(0)$ can also be expressed as
\begin{equation}
\label{static2} \chi_{\sc reg}(0) = \int_{-\infty}^{\infty}
\frac{d\omega}{\pi}\  \frac{\chi_{\sc reg}''(\omega)}{\omega} \;.
\end{equation}
Equations~(\ref{imchi3}) and (\ref{static2}) are compatible only if
\begin{equation}
\label{imchilimit2} \lim_{M/\gamma_0 \to \infty} {\chi_{\sc
reg}''(\omega) \over \omega} = \frac{\pi}{\kappa_p}\
\delta(\omega).
\end{equation}
Therefore, the system has no intrinsic dynamics in this limit.
In terms of the original spin model
this is a simple consequence of the form of our starting Hamiltonian,
Eq.~(\ref{hamisb}): if $\Delta = 0$ the spin variables
commute with the Hamiltonian and are thus constants of the motion.
In terms of the particle interpretation the limit $M/\gamma_0\to\infty$
corresponds to an infinitely massive particle that is not able to move
or to the limit of zero friction where there is no dissipation.

The expressions in Eq.~(\ref{c_no_inertia_asymptotic}) can be
readily generalized to non-Ohmic baths. We find that in the
long-time limit:
\begin{eqnarray}
\label{c_asymptotic_nonOhomic} C(t) - q_{\sc ea}\;\propto\;\left\{
\begin{array}{ll} \cos\left({s + 2 \over 4}\;\pi\right)\;t^{- \left(1 + {s\over 2}\right)} \;, & T = 0\;,\\
\\ \cos\left({s \over 4} \pi \right)\; t^{- {s \over 2}} \;, & T \gg \hbar/t\;.
\end{array}
\right.
\end{eqnarray}

\section{Real-time dynamics}
\label{real-time2}
\setcounter{equation}{0}
\renewcommand{\theequation}{\thesection.\arabic{equation}}

In this Section we study the real-time dynamics of the
system coupled to the environment. We use the dynamic equations
for the symmetrized correlation and linear response functions
derived in \cite{Culo} with the Schwinger-Keldysh formalism
and we solve them numerically, as a function of time,
for different couplings to the bath and different environments.
We compare the results to the ones obtained in the previous
Section with the imaginary time formalism.

\subsection{The dynamic equations}

The dynamic equations for the model defined in Section~2
were derived in \cite{Culo}.
They are of the Schwinger-Dyson form and read
\begin{eqnarray}
(M\partial^2_t + z(t) )R(t,t')
&=& \delta(t-t') + \int_0^\infty dt'' \, \Sigma(t,t'') R(t'',t')
\; ,
\label{schwingerR}
\\
(M\partial^2_t + z(t)) C(t,t')
&=& \int_{0}^\infty dt'' \, \Sigma(t,t'') C(t'',t') + \int_{0}^{t'} dt'' \,
D(t,t'') R(t',t'')
\; ,
\label{schwingerC}
\end{eqnarray}
with the equal-times conditions
$C(t,t) = 1$ and
$R(t,t)=0$
and
\begin{eqnarray}
\lim_{t'\to t^-} \partial_t R(t,t') &=& \frac{1}{M}
\; ,
\nonumber\\
\lim_{t'\to t^+} \partial_t R(t,t') &=& 0
\; ,
\\
\lim_{t'\to t^-} \partial_t C(t,t') &=&
\lim_{t'\to t^+} \partial_t C(t,t') = 0
\; .
\end{eqnarray}
The equation for the Lagrange multiplier, $z(t)$,  reads
\begin{eqnarray}
z(t)
&=&
\int_0^t dt'' \left[
\Sigma(t,t'') C(t,t'') + D(t,t'') R(t,t'')
\right]
\nonumber\\
& &
+ M \int_0^t dt'' \int_0^t  dt''' \,
 (\partial_t R(t,t'') ) \,D(t'',t''')\, (\partial_t R(t,t''') )
\\
\nonumber & & + \left. M^2 \left[ \partial_t R(t,s) \partial^2_{s t} C(s,t) -
\partial^2_{s t} R(t,s) \partial_{t'} C(s,t')
\right]\right|_{\begin{array}{l}
s \to 0\\
t \to t'
\end{array}}
\end{eqnarray}
The total self-energy and vertex include the interaction with the
bath and are given by
\begin{eqnarray}
\Sigma(t,t')
&\equiv&
- 4 \eta(t-t')
-\frac{p \tilde{J}^2}{\hbar}
\mbox{Im} \left[ C(t,t')-\frac{i \hbar}{2} R(t,t') \right]^{p-1}
\; ,
\label{sigma2}
\\
D(t,t')
&\equiv&
2 \hbar \nu(t-t') +
\frac{p \tilde{J}^2}{2}
\mbox{Re}\left[ C(t,t')-\frac{i\hbar}{2}( R(t,t')+ R(t',t)) \right]^{p-1}
\;
\label{D}
\end{eqnarray}
with
\begin{eqnarray}
\nu(t-t') &=&  \int_0^\infty d\omega I(\omega) \,
\coth\left( \frac12 \beta \hbar \omega \right) \;
\cos(\omega (t-t'))
\; ,
\label{nu}
\\
\eta(t-t') &=&
-\theta(t-t') \,
\int_0^\infty d\omega \; I(\omega) \, \sin(\omega (t-t'))
\; .
\label{eta}
\end{eqnarray}
The spectral density of the bath, $I(\omega)$,
has been defined in Eq.~(\ref{Iomega})~\cite{comment}.

In the following we shall compare the effect of
environments with different values of $s$ and 
using different coupling strengths. The high-frequency
cut-off, $\omega_c$, is introduced to avoid the divergence of
$\nu(\tau)$. In the Sub-Ohmic case, when we solve the equations
numerically,  we also need a low frequency cut-off, that we
impose in a hard manner by including a factor $\theta(\omega-b)$
in the definition of $I(\omega)$.

The kernels $\nu$ and $\eta$ can be computed for
all values of $s$. In the numerical solution to
the set of coupled integro-differential equations
(\ref{schwingerR})-(\ref{schwingerC})
it is more useful to use the integral of the
kernel $\eta$,
$\hat\eta(\tau)\equiv \int_\tau d\tau' \eta(\tau')$, that reads
\begin{eqnarray}
\hat\eta(\tau)
&=&
\frac{\alpha \hbar}{2 \pi} \;
\frac{\omega_c}{\left(1+\omega_c^2 \tau^2\right)^{s/2}}
\;
\cos\left( s \mbox{Arctg}(\omega_c \tau) \right) \Gamma(s)
\end{eqnarray}
and, when $s$ takes the values $1/2,1,3/2$, it becomes
\begin{eqnarray}
\hat\eta(\tau) &\sim&
\displaystyle{
\left\{
\begin{array}{rcl}
\sqrt{\frac{\omega_c}{\tau}}
& s=1/2 & \;\;\; {\rm subOhmic}
\nonumber\\
\frac{1}{\tau}
& s=1 & \;\;\; {\rm Ohmic}
\nonumber\\
\frac{1}{\sqrt{\omega_c \tau^3}}
& s=3/2 & \;\;\; {\rm superOhmic}
\end{array}
\right.
}
\end{eqnarray}
On the right-hand-side
we have written the limiting form for $\omega_c\tau\gg 1$.
It is clear that, as for the imaginary-time kernels,
the dependence on $\omega_c$ is very different in each
of these cases.

We shall rescale  the real time and the  other parameters
and functions in the dynamic equations to match the definitions that we
used in the
Matsubara calculation. Under the rescaling of time, $t \to
\tilde{J}/\hbar\;t$, the correlation function remains
unchanged and the response transforms as $R \to \hbar R$. 
The rescaled dynamic equations are identical to
Eqs.~(\ref{schwingerR}) and (\ref{schwingerC}) with $M$ replaced by
$\Gamma^{-1}$.

\subsection{Numerical study of the real-time dynamics}

As shown in Section~\ref{sec:results},
both static and dynamic transition lines
depend strongly on the strength of
the coupling between system and bath.
We can also see this effect by following
the real-time dynamics of the system coupled to the environment.
We have solved Eqs.~(\ref{schwingerR})-(\ref{schwingerC})
numerically with a predictor-corrector algorithm that allows
us to reach long times with a high accuracy.
For each set of parameters we have checked the data collapse
for different values of the iteration step $h$ in the
discretized equations. In general, there is a good
collapse for $h\leq 0.02$ and, typically, we have used
$h= 0.01$ and  $h=0.02$.

\subsubsection{Effect of the interactions: localization against
glassy behavior}

\begin{figure}
\epsfxsize=5in
\centerline{\epsffile{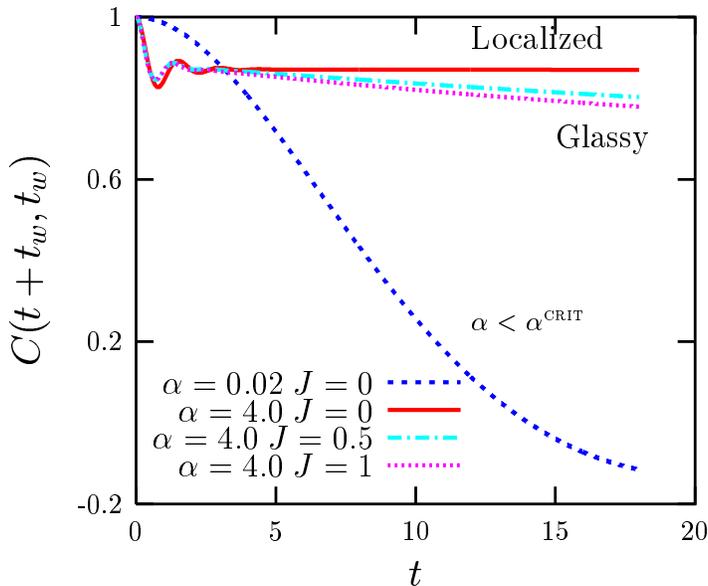}}
\vspace{-.2cm}
\caption{The decay of the symmetrized correlation in three cases:
localization for $\alpha=4$ and $\tilde J=0$, glassy decay for two
nonvanishing values of $\tilde J$, $\tilde J=0.5$ and $\tilde J=1$ and 
a simple decay towards zero for the case of a small coupling to the bath, 
$\alpha=0.2$ and $\tilde J=0$. We have chosen a subOhmic bath with 
$s=0.5$, $\omega_{ph}=5$ and $\omega_c=10$. The quantum parameter 
$\Gamma$ equals one.}
\label{local1}
\end{figure}

In the Introduction and Section~1 we recalled several results for 
localization in dilute two-level systems coupled to a bath. 
In this paper we focus on 
a soft spin version of the interacting problem. Our first aim 
is to determine the 
effect of the coupling $\tilde J$ on the localization properties 
of this system from a real-time dynamic point of view. 
In Fig.~\ref{local1} we show the decay of the symmetrized 
correlation $C(t+t_w,t_w)$ using a subOhmic bath with $s=0.5$, 
$\omega_c=10$ and $\omega_{ph}=5$. 
The three upper curves were obtained 
for $\alpha=4$ and changing the value of the {\sc sg} coupling
strength $\tilde J$. When $\tilde J=0$ the system localizes for
$\alpha>\alpha^{\sc crit}$: for any 
$t_w$ and long enough $t$ the correlation reaches a plateau and it does not 
decay below this value. 
When a small coupling is switched on the 
decay changes. The correlation approaches a plateau for small values of 
$t-t_w$ but it subsequently quits the plateau and decays towards zero. 
The system has glassy non equilibrium dynamics that we shall quantify below.
Finally, when the coupling to the bath is very small and $\tilde J=0$ the 
system does not localize and the correlation decays to zero with wide 
oscillations.

\subsubsection{Dynamics in the paramagnetic phase}

For a chosen coupling to a bath, 
at sufficiently high values of $\Gamma$ and/or $T$ the system 
equilibrates with the environment and it quickly reaches a stationary regime where the
quantum fluctuation - dissipation theorem  ({\sc fdt})
is satisfied. This property has been proven for the $p$ 
spin model in \cite{Culo},  for the large $N$ 
fully connected Heisenberg SU(N) model in  \cite{Bipa} and 
for a soft version of the
quantum model in \cite{Chamon}. In all cases the systems were 
coupled to an Ohmic environment and the limit of 
weak coupling, $\lim_{\alpha\to 0} \lim_{t_w\to\infty}$, 
was considered.
The correlation and response
have a rapid decay towards zero with oscillations that depend on
the value of the quantum parameter $\Gamma$ and, as we show here,
on the coupling to and the type of environment used.

\begin{figure}
\epsfxsize=6.5in
\centerline{\epsffile{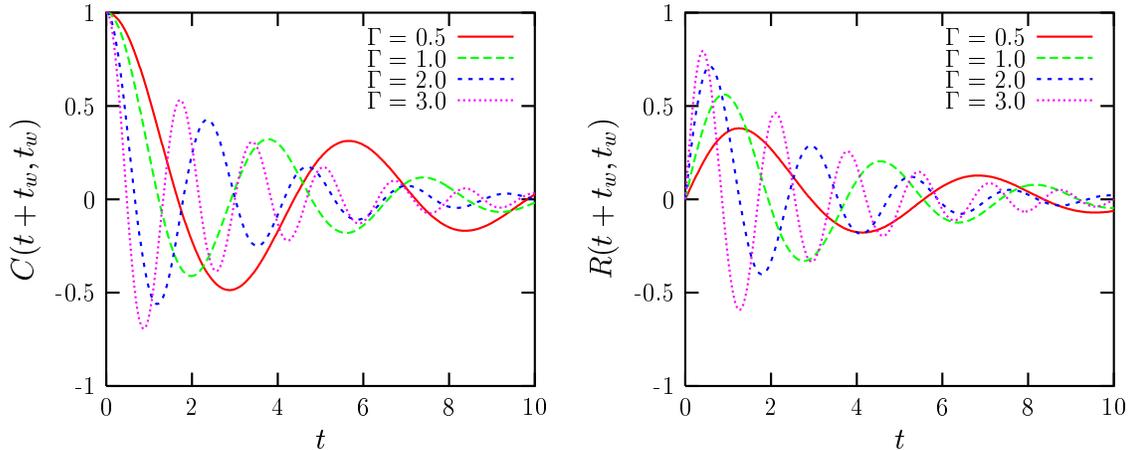}}
\vspace{-.2cm}
\caption{The stationary autocorrelation (left) and response (right)
functions 
at $T=2$
for several values of the
quantum parameter $\Gamma$, given in the key. The bath is Ohmic and
$\omega_c=5$, $\alpha=0.8$.}
\label{corrT2_label}
\end{figure}

\begin{figure}
\epsfxsize=6.5in
\centerline{\epsffile{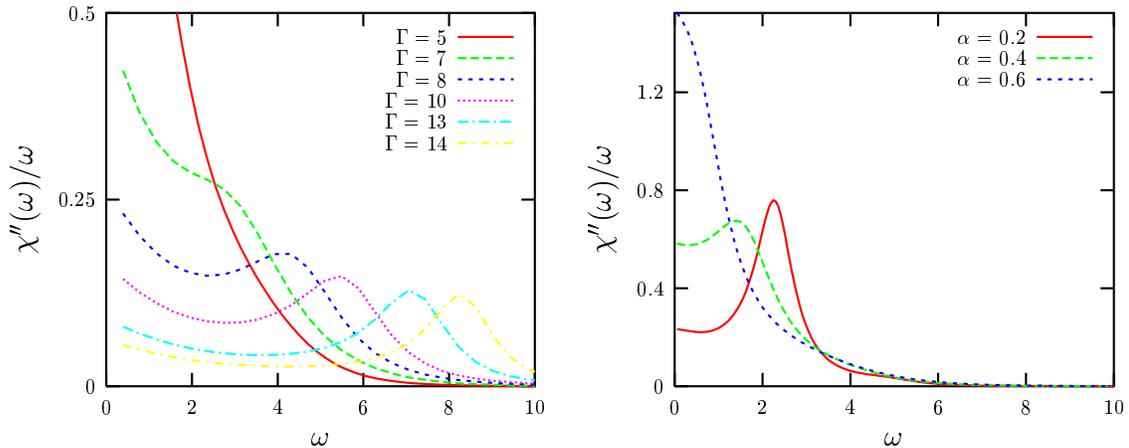}}
\caption{The frequency dependence of $\chi''(\omega)/\omega$.
Left panel: $T=0$, $\alpha=0.6$, $\omega_c=5$
 and several values of $\Gamma$. Right
panel: $\Gamma=5$, $T=0$, $\omega_c=10$
and several values of $\alpha$. The environment is Ohmic in both cases.}
\label{chisec}
\end{figure}

In this Section we analyze the effect of the quantum fluctuations and the 
bath on the conclusions mentioned above. 
We first consider a fixed Ohmic environment, {\it i.e.}
we take $s=1$ and we fix $\omega_c=10$.
We display in Fig.~\ref{corrT2_label} the decay of the correlation and response
functions for different values of $\Gamma$ in the {\sc pm} phase.
It is clear from the figure that
the period of the oscillations
decreases with $\Gamma$. In order to quantify this dependence one can
compute $\chi''(\omega)/\omega$ and follow the evolution of the
peaks. We show two examples in Fig.~\ref{chisec}. The data on the left panel
correspond to those on 
Fig.~\ref{corrT2_label}. On the right panel we represent
$\chi''(\omega)/\omega$ for $T=0$, $\Gamma =5$ and several values of
$\alpha$. 
For small values of $\alpha$ the system is deep in the {\sc pm} phase
and there is a well defined peak in $\chi''(\omega)/\omega$ at a finite
value  $\omega_0$ that increases with decreasing $\alpha$.
At high enough values of $\alpha$ a tail at low frequencies starts
developing, indicating that the dynamics is slower and that
the system approaches the transition towards the glassy
phase. Eventually, as discussed in
Section~\ref{formalism}, for high enough $\alpha$, the parameters
fall below the transition and the system becomes glassy
with slow dynamics.

\subsubsection{Dynamics in the glassy phase}

\begin{figure}
\epsfxsize=6.5in
\centerline{\epsffile{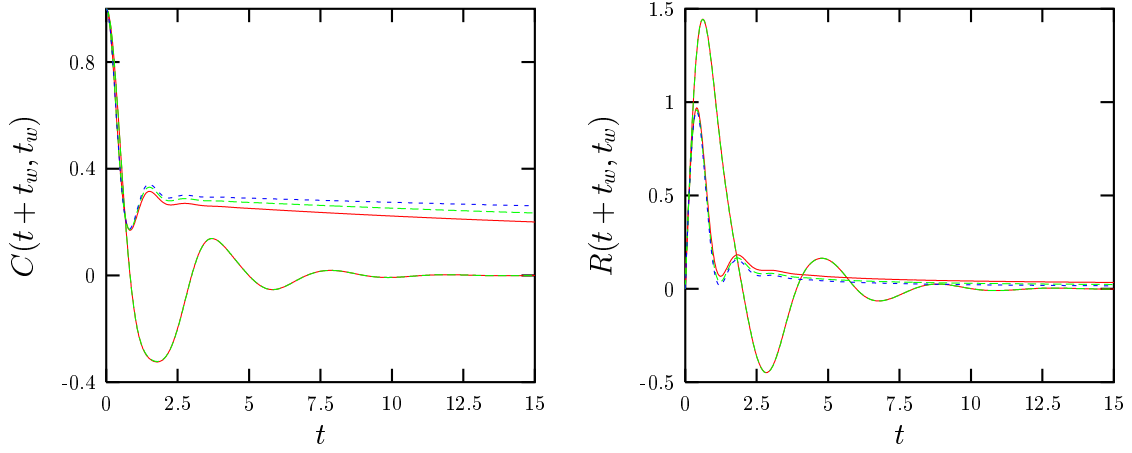}}
\caption{Left: The auto-correlation $C(t+t_w,t_w)$ as a function of $t$
for  $\alpha=0.2$ ({\sc pm} phase) and $\alpha=1$ ({\sc sg} phase).
The temperature is $T=0.1$ in both cases.
The different curves correspond
to different waiting times, $t_w=5,10,20$.  For $\alpha=0.2$
the curves collapse on an asymptotic one, while for $\alpha=1$ they
show  aging effects. Right: The
response $R(t+t_w,t_w)$  as a function of $t$
for the same parameters. The effect is similar.}
\label{compCR}
\end{figure}

In Figs.~\ref{compCR}
we compare the behavior of the correlation and response functions
in the glassy phase 
when the system is coupled to an Ohmic environment through different
coupling constants. We choose $T=0.1$, $\Gamma=4$ and we compare the
effect of $\alpha=0.2$ and $\alpha=1$. The high-frequency cut-off is
$\omega_c=5$.
From the discussion in Section 3 we expect that  the system is 
in the {\sc pm} phase in the first
case and  in the {\sc sg} phase in the second.  
This is seen in Fig.~\ref{compCR}.
For $\alpha=0.2$ the correlations rapidly reach a
stationary regime and they oscillate around zero. For
$\alpha=1$ the behavior is different.
There is a first rapid
decay towards a plateau, that has a low value,
and then, a slow and
monotonic decay towards zero. Aging effects are
apparent from the figure. The response function also
shows a qualitatively
different behavior according to the value of $\alpha$. In one case
it quickly acquires a stationary oscillatory behavior around zero,
in the other it has a long tail as expected in a glassy system.
We then conclude that the system has undergone a
dynamic
phase transition between the {\sc pm} and {\sc sg} phases
at an intermediate value of $\alpha$.

An approximate expression for the
dependence of the Edwards-Anderson parameter on $\Gamma$ and
$\alpha$, at $T\sim 0$,
has been obtained in Section~\ref{sec:lowfreq}. We can also check this
law by estimating the value of $q_{\sc ea}$ from the numerical
solution of the real-time equations. If
we plot  the correlation function for several values of
$\alpha$ in a log-log scale the plateau at $q_{\sc ea}$
can be easily identified. It is a slowly growing function
of $\alpha$ that is rather well described  by 
Eq.~(\ref{qeanear1ohmic}).

\begin{figure}
\epsfxsize=6.5in
\centerline{\epsffile{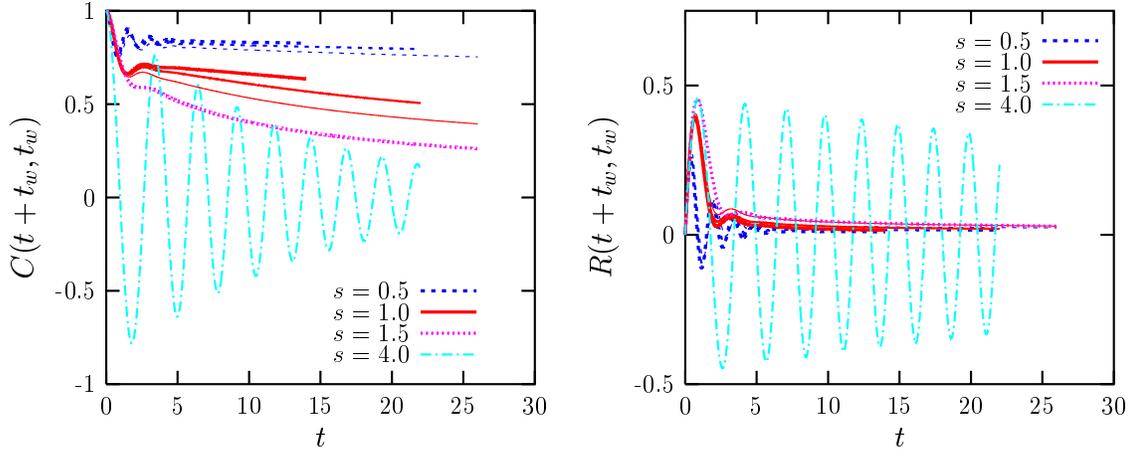}}
\caption{The symmetrized autocorrelation and the linear response
 at $T=0.1$,  for $\Gamma=1$.
We compare the effect of a sub-Ohmic ($s=0.5$), an Ohmic
($s=1$) and a two super-Ohmic ($s=1.5$, $s=4$) baths.
See the key for the details.
The coupling to the bath is kept fixed to $\alpha=3$ and the
high-frequency cut-off equals the phonon frequency, 
$\omega_c= \omega_{ph}=5$. }
\label{compCR_aaG1T01g1_paper}
\end{figure}

We also investigated the effect of different environments (different
$s$) of the same strength (same $\alpha$) using the same value of the 
high-frequency cut-off that we took equal to $\omega_{ph}$.
From the discussion in Section 3.3 for some values of $\omega_{ph}$ 
we expect the relaxation to be  
slowest for the subOhmic bath, intermediate in the  Ohmic case 
and faster for a superOhmic environment. This is illustrated  in
Fig.~\ref{compCR_aaG1T01g1_paper}. The decay
is slower when $s=0.5$ than in the other cases. In the
extreme case of $s=4$ the system has gone across the
transition towards the {\sc pm} phase. However, this behavior is not
generic. 

The relation between the correlation and response plays a key role in the
description of the dynamic behavior of glassy systems. When the system is
in equilibrium, this relation is model independent and it is given by the
fluctuation-dissipation theorem ({\sc fdt}).
When the system is glassy and it evolves 
out of equilibrium, the conditions to prove the theorem are not
satisfied but simple generalizations have been exhibited in
a number of models~\cite{Cuku,Culo}. 

The quantum {\sc fdt} for a system in equilibrium, in the rescaled variables, reads:
\begin{equation}
R(t) =
\theta(t) \; i \int_{-\infty}^{\infty}  d\omega \; e^{-i \omega t}
\;
\tanh\left( \frac{\beta\tilde J \omega}{2}\right)
\;
{\tilde C}(\omega)
\label{FDT3}
\end{equation}
where
\begin{equation}
{\tilde C}(\tilde \omega)
=
2 \mbox{Re} \int_{0}^{\infty}  d\tau \; e^{i\omega t} \; C(\tau)
\label{FDT4}
\; .
\end{equation}
The quantum {\sc fdt} is 
an integral relation between the stationary linear response and the 
symmetrized correlation function. 

The asymptotic dynamics in the glassy phase take place in two time scales that are 
separated by the plateau in the correlation function. As shown in \cite{Culo}
for the weak coupling limit, the stationary part of the decay, when the 
correlation decays from $1$ to $q_{\sc ea}$ is such that the quantum {\sc fdt} 
holds. We have checked that this result also holds when the system is 
strongly coupled to a nonOhmic bath. In the weak coupling limit, 
when the correlation decays beyond ${\sc ea}$, the relation between linear
response and correlations 
takes the form of the classical {\sc fdt} and it reads
\begin{equation}
 R(t) = \theta(t) \; \beta_{\sc eff}\tilde{J} \;
\frac{\partial}{\partial t} C(t)
\; .
\end{equation} 
where $\beta_{\sc eff}$ is the inverse 
of an effective temperature~\cite{Cukupe} $T_{\sc eff}$
and $\tilde J$ appears since we have rescaled time.
A concrete way of testing the validity of this equation is to 
plot the integrated response function 
\begin{equation}
\chi(t+t_w,t_w) \equiv \int_{t_w}^{t+t_w} dt' R(t+t_w,t') 
\end{equation}
against the symmetrized correlation $C(t+t_w,t_w)$ for a long
enough $t_w$, and using $t$ as a parameter. For short time-differences, 
when $t-t_w \ll t_w$ and $C(t+t_w,t_w)> q_{\sc ea}$, this construction does 
not have any particular meaning and the curve is nonmonotonic with strong 
oscillations. Instead, when $t-t_w \approx t_w$ or longer
and $C(t+t_w,t_w)< q_{\sc ea}$,
the curve becomes a straight line of slope $-1/T_{\sc eff}$. 

\begin{figure}
\epsfxsize=6.5in
\centerline{\epsffile{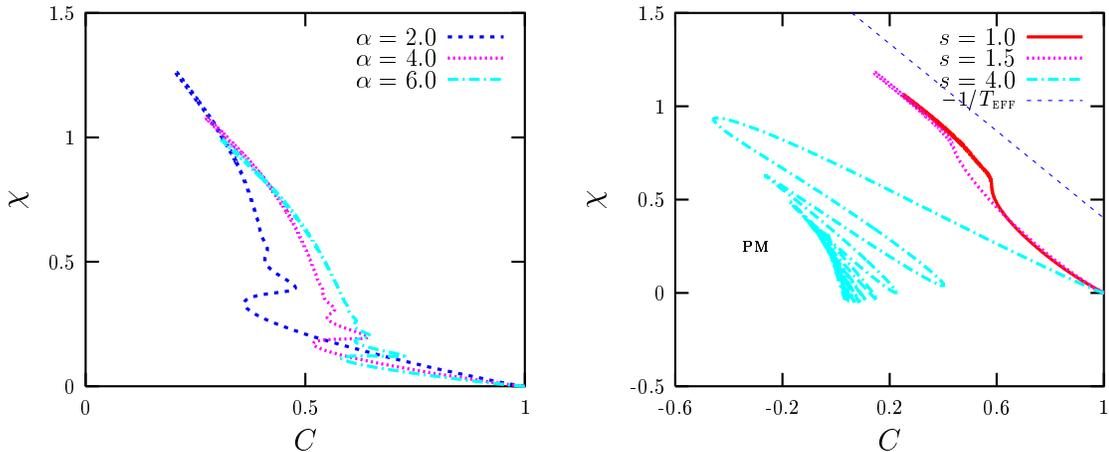}}
\caption{Left: The parametric
$\chi$ against $C$ curves for different values of
the coupling to the environment, at fixed $T=0.1$, $\omega_c=5$ and
using an Ohmic bath. Right: the dependence of the $\chi$ against
$C$ plot on the kind of bath used. We include a straight line as a guide to the eye.
In all curves $T=0.1$, $\alpha=2$ and $\omega_c=5$
for all curves.}
\label{compchi}
\end{figure}

In Figs.~\ref{compchi} we display the $\chi$ against $C$ plots
for different values of the parameters, explained in the caption and
keys. The panel on the left shows the $\chi$ {\it vs} $C$ curve
for different values of the coupling $\alpha$. The slopes of the
curves, and hence $T_{\sc eff}$,
change smoothly and there is a clear non trivial dependence on this
parameter. The panel on the right displays the $\chi(C)$ plot 
for a fixed value of the coupling $\alpha$ and several values of
$s$, $s=1,1.5,4$.  The first two cases are in the glassy phase while the 
latter falls in the {\sc pm} phase and the parametric plot does not 
show a straight line piece. It is difficult to decide from these 
figures if the slopes depend on
$s$ or not.  

In order to sharpen our conclusions about the dependence of $T_{\sc eff}$ on 
the characteristics of the environment we take profit of the empiric relation
between $T_{\sc eff}$ and the breaking point parameter $m$ in the replica analysis
of the same model, $T_{\sc eff}=T/m$. In Section~3 we developed 
a low temperature, low frequency 
approximation to solve the saddle point equations stemming from the replicated 
Matsubara analysis of this problem. In these limits we derived a set of 
equations that link $T/m$ to $\alpha$ and $s$ 
that can be solved numerically.  We found that  
for fixed $s$ the effective temperature $T_{\sc eff}$ 
is a growing function of $\alpha$. This result is reminiscent to 
the dependence of $T_{\sc eff}$ on the external temperature $T$ 
in a classical problem: the lower $T$, the higher $T_{\sc eff}$ meaning that 
higher values of the effective temperature are reached when the system is 
deeper in the glassy phase. Each curve approaches one 
when $\alpha\to \infty$ and the 
corrections can be read from the asymptotic analysis presented in Section~3. 
The dependence of $T_{\sc eff}$ on $s$ is weak but non-monotonic. 
(We have already encountered a non-monotonicity related to 
the fact that the factor $\omega_{ph}^{1-s}$ changes the coupling between 
system and bath differently for different values of $s$.)

\section{Conclusions}
\label{conclusions}

In this article we discussed the effect of a 
quantum environment on the nonequilibrium dynamic properties 
of an interacting quantum glassy system. We have shown that, as in the 
case of a simple {\sc tls}, the influence of the quantum bath 
is very important. 

Two limits of the quantum model are easy to 
derive or were already known. 
On the one hand, in the absence of interactions, the calculations
shown in Section~2 and the numerical results of Section~4.1.1 prove 
that when the model is coupled to a subOhmic bath, it 
undergoes a localization transition  at a critical value of the coupling
$\alpha$. The localized phase is characterized by a symmetrized two-time 
correlation function that, as a function of time-difference, 
approaches a non-vanishing asymptotic value and {\it never} decays 
to zero. On the other hand, it was known that when interactions are 
switched on and 
the limit of weak coupling, $\alpha\to 0$, is taken, the model has glassy 
dynamics with a symmetrized correlation function that depends on the 
waiting-time and decays in two-steps with a first approach to a plateau 
and a second decay towards zero~\cite{Culo}. 

The aim of this article was to analyze the combined effects of the 
interaction ($\tilde J\neq 0$) and a strong coupling ($\alpha\neq 0$)  
to quantum environments of different types (different $s$). 
We summarize our findings as follows:

First, we determined if the model has a localized phase 
in the presence of interactions. How to define such a phase for an 
interacting system is a difficult question (see, {\it e.g.}, \cite{Vojta}).
Here, we adopted as evidence for a localized phase the fact that 
for a long enough waiting-time $t_w$ 
the correlation function does not decay to zero at any time-difference
$t-t_w$. With this criterium we saw that, as expected, 
there is no localized phase
when interactions are switched on. 

This result can be interpreted {\it a posteriori} by resorting to the 
concept of effective temperatures generated by the nonequilibrum
dynamics of glassy systems. Indeed, it has been shown for classical systems 
that the modification of the fluctuation-dissipation theorem observed 
in systems evolving slowly out of equilibrium is related to 
the self-generation of effective temperatures (typically higher than the 
one of the environment)~\cite{Cukupe}. The proof presented in \cite{Cukupe}
has not been extended to quantum systems yet. However, as argued in \cite{Culo}
for the quantum model studied in this paper when weakly coupled to 
an environment, the slow part of the 
relaxation looks {\it classical}, with a quantum fluctuation -- 
dissipation relation that became classical with  
an effective temperature
that is higher than the temperature of the environment. 
In particular, when the model is coupled a quantum bath at zero 
temperature it acquires a non-vanishing effective temperature.  
This effect has been observed 
in other quantum glassy systems too~\cite{Chamon,Bipa}.
When the system is strongly coupled to the environment
the relaxation slows down with respect to the weakly coupled case. 
However, the two step relaxation remains with a slow regime controlled by a
nonvanishing effective temperature.
Thus, we conclude that the generation of an effective temperature 
by the interactions 
is consistent with the fact that the system does not localize. 
It is well known that even in simple {\sc tls} the localization effects  
disappear at finite temperature. 

Next, we analyzed the effect of a strong coupling to an environment
on the glassy properties of the model. We showed that stronger couplings 
to the bath favor the glassy phase for any type of bath. By this we mean 
that for larger value of $\alpha$ the area  of the spin-glass phase 
on the $(T, \Gamma)$ plane increases. We also characterized the 
dependence on $\alpha$ of several properties of the system as the 
Edwards-Anderson parameter, the effective temperature, etc. 

Finally, we studied the effect of different types of baths. Concerning
this issue the conclusions are cumbersome given the fact that a new 
parameter, the phonon frequency $\omega_{ph}$, appears in the spectral 
density when $s\neq 1$. If $\omega_{ph}$ is not equal to 
one, the effect of different baths are complicated. 
For instance, the dependence of $q_{\sc ea}$ on $s$ can be nonmonotonic
as well as the location of the critical line on the 
$(T,\Gamma)$ plane.
We exhibited some examples but we cannot draw general conclusions concerning
this issue.

The motivation for this study were manifold. 
The effect of quantum environments on interacting macroscopic quantum systems
is a problem that is now being revisited in the context of quantum
computing~\cite{Juanpa}. Decoherence, or how quantum 
interference effects are lost due to the interaction with the environment, 
has to be as much reduced as possible to make a quantum computer 
performant. Again in the context of quantum computing, an isolated 
Edwards-Anderson quantum model in a random transverse field has 
been proposed to mimic an isolated  
quantum computer with (short-range) interactions between the spins 
(that represent qubits) and with static ``imperfections'' in the individual
two-level system energies~\cite{Toulouse}. In this work we analyzed
a soft limit of a disordered quantum model with long-range 
$p>2$ interactions in a
transverse field. It would be very interesting to see which, if any, of 
our conclusions are modified if the soft spin limit is lifted and, 
even more importantly, if a finite dimensional model is considered. 
This project, however, is a very difficult one. 

On a more physical side, glassy phases at very low temperatures 
where quantum fluctuations are important have been identified in a 
number of physical systems. 
In the proper analysis of these systems the role played 
by the quantum environment has to be taken into account. Our results 
are a first step towards the characterization of the effects of the 
environment. Again, it would be interesting to go beyond the mean-field 
limit and derive similar results for a finite dimensional model.

\vspace{2cm}

\noindent\underline
{Acknowledgments}
We especially thank L. Ioffe for suggesting this problem to us and 
for discussions at the initial stages of this work. 
We thank a travel grant from the international collaboration ECOS-Sud and  
the research program ``Probl{\`e}mes d'optimisation et syst{\`e}mes 
d{\'e}sordonn{\'e}s quantiques'' (ACI Jeunes Chercheurs) for financial support. 
G. L. is associated with CONICET Argentina and he 
acknowledges Fundaci{\'o}n Antorchas for financial support.
L. F. C. is research associate at ICTP, Trieste, Italy.
After the 1st October 2001  the address of C. da S. S. 
is Instituto Carlos I de Fisica Teorica y Computational, 
Universidad de Granada, E-18071 Granada Spain. He 
acknowledges financial support from the Portuguese Research Council 
under grants PRAXIS XXI/BPD/16303/98 and SFRH/BPD/5557/2001.

\appendix

\pagebreak


\begin{thebibliography}{99}
\bibitem{comment} In Ref.~\cite{Culo} the spectral density
was defined as $4\overline M \gamma_0/\pi (\omega/\Lambda)^{s-1}
\omega \exp(-\omega/Lambda)$. The relation between the constants
$\overline M \gamma_0$ and $\alpha$ is then $2\overline M \gamma_0=
\alpha\hbar$.
\end{thebibliography}
\end{document}